\documentclass[twoside,twocolumn,9pt]{article}
\usepackage{extsizes}
\usepackage[super,sort&compress,comma]{natbib} 
\usepackage[version=3]{mhchem}
\usepackage[left=1.5cm, right=1.5cm, top=1.785cm, bottom=2.0cm]{geometry}
\usepackage{balance}
\usepackage{mathptmx}
\usepackage{sectsty}
\usepackage{graphicx} 
\usepackage{lastpage}
\usepackage[format=plain,justification=justified,singlelinecheck=false,font={stretch=1.125,small,sf},labelfont=bf,labelsep=space]{caption}
\usepackage{float}
\usepackage{fancyhdr}
\usepackage{fnpos}
\usepackage[english]{babel}
\addto{\captionsenglish}{%

}
\usepackage{array}
\usepackage{droidsans}
\usepackage{charter}
\usepackage[T1]{fontenc}
\usepackage[usenames,dvipsnames]{xcolor}
\usepackage{setspace}
\usepackage[compact]{titlesec}
\usepackage{hyperref}
\usepackage{epstopdf}%This line makes .eps figures into .pdf - please comment out if not required.

\definecolor{cream}{RGB}{222,217,201}

\usepackage{bm}
\usepackage{comment}
\usepackage{amssymb,amsfonts,amsmath}

\begin{document}

\pagestyle{fancy}
\thispagestyle{plain}
\fancypagestyle{plain}{
%%%HEADER%%%
\renewcommand{\headrulewidth}{0pt}
}
%%%END OF HEADER%%%

%%%%%%%%%%%%%%%%CUSTOM COMMANDS %%%%%%%%%%%%%%%%%%%
\newcommand*{\fznote}[1]{{\color{blue}({\bf \footnotesize FZ: } #1)}}
\newcommand*{\iknote}[1]{{\color{magenta}({\bf \footnotesize IK: } #1)}}
\newcommand{\matr}[1]{\boldsymbol{\bm{#1}}}
\renewcommand{\vec}{\bm}
\newcommand*{\rev}[1]{{\color{black}#1}}

%%%PAGE SETUP - Please do not change any commands within this section%%%
\makeFNbottom
\makeatletter
\renewcommand\LARGE{\@setfontsize\LARGE{15pt}{17}}
\renewcommand\Large{\@setfontsize\Large{12pt}{14}}
\renewcommand\large{\@setfontsize\large{10pt}{12}}
\renewcommand\footnotesize{\@setfontsize\footnotesize{7pt}{10}}
\makeatother

\renewcommand{\thefootnote}{\fnsymbol{footnote}}
\renewcommand\footnoterule{\vspace*{1pt}% 
\color{cream}\hrule width 3.5in height 0.4pt \color{black}\vspace*{5pt}} 
\setcounter{secnumdepth}{5}

\makeatletter 
\renewcommand\@biblabel[1]{#1}            
\renewcommand\@makefntext[1]% 
{\noindent\makebox[0pt][r]{\@thefnmark\,}#1}
\makeatother 
\renewcommand{\figurename}{\small{Fig.}~}
\sectionfont{\sffamily\Large}
\subsectionfont{\normalsize}
\subsubsectionfont{\bf}
\setstretch{1.125} %In particular, please do not alter this line.
\setlength{\skip\footins}{0.8cm}
\setlength{\footnotesep}{0.25cm}
\setlength{\jot}{10pt}
\titlespacing*{\section}{0pt}{4pt}{4pt}
\titlespacing*{\subsection}{0pt}{15pt}{1pt}
%%%END OF PAGE SETUP%%%

%%%FOOTER%%%
\fancyfoot{}
\fancyfoot[LO,RE]{\vspace{-7.1pt}\includegraphics[height=9pt]{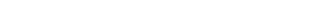}}
\fancyfoot[CO]{\vspace{-7.1pt}\hspace{13.2cm}\includegraphics{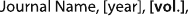}}
\fancyfoot[CE]{\vspace{-7.2pt}\hspace{-14.2cm}\includegraphics{head_foot/RF}}
\fancyfoot[RO]{\footnotesize{\sffamily{1--\pageref{LastPage} ~\textbar  \hspace{2pt}\thepage}}}
\fancyfoot[LE]{\footnotesize{\sffamily{\thepage~\textbar\hspace{3.45cm} 1--\pageref{LastPage}}}}
\fancyhead{}
\renewcommand{\headrulewidth}{0pt} 
\renewcommand{\footrulewidth}{0pt}
\setlength{\arrayrulewidth}{1pt}
\setlength{\columnsep}{6.5mm}
\setlength\bibsep{1pt}
%%%END OF FOOTER%%%

%%%FIGURE SETUP - please do not change any commands within this section%%%
\makeatletter 
\newlength{\figrulesep} 
\setlength{\figrulesep}{0.5\textfloatsep} 

\newcommand{\topfigrule}{\vspace*{-1pt}% 
\noindent{\color{cream}\rule[-\figrulesep]{\columnwidth}{1.5pt}} }

\newcommand{\botfigrule}{\vspace*{-2pt}% 
\noindent{\color{cream}\rule[\figrulesep]{\columnwidth}{1.5pt}} }

\newcommand{\dblfigrule}{\vspace*{-1pt}% 
\noindent{\color{cream}\rule[-\figrulesep]{\textwidth}{1.5pt}} }

\makeatother
%%%END OF FIGURE SETUP%%%

%%%TITLE, AUTHORS AND ABSTRACT%%%
\twocolumn[
  \begin{@twocolumnfalse}
{\includegraphics[height=30pt]{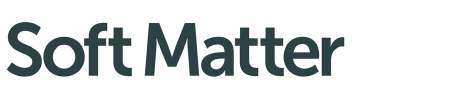}\hfill\raisebox{0pt}[0pt][0pt]{\includegraphics[height=55pt]{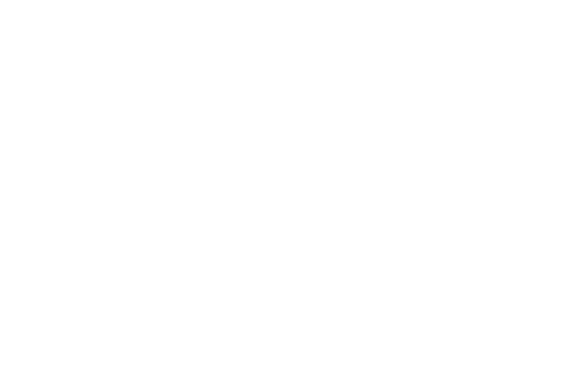}}\\[1ex]
\includegraphics[width=18.5cm]{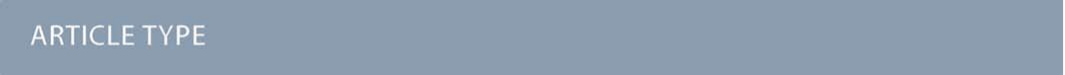}}\par
\vspace{1em}
\sffamily
\begin{tabular}{m{4.5cm} p{13.5cm} }

\includegraphics{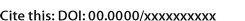} & \noindent\LARGE{\textbf{Reshaping and Enzymatic Activity may allow Viruses to move through the Mucus}} \\%Article title goes here instead of the text "This is the title"
\vspace{0.3cm} & \vspace{0.3cm} \\

 & \noindent\large{Falko Ziebert,\textit{$^{a,b}$} Kenan G. Dokonon\textit{$^{c}$} 
 and Igor M. Kuli\'{c}\textit{$^{d,e}$}} 
 \\%Author names go here instead of "Full name", etc.

\includegraphics{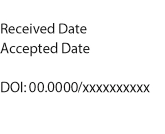} & \noindent\normalsize{Filamentous viruses 
like influenza and torovirus 
often display systematic bends and arcs of mysterious physical origin. 
We propose that such  viruses  undergo an instability 
from a cylindrically symmetric to a toroidally curved state.   
This ``toro-elastic'' state emerges via spontaneous symmetry breaking 
under prestress due to short range spike protein interactions 
magnified by %the filament's 
surface topography.
Once surface stresses are sufficiently large, 
the filament buckles and the %toroidal, 
curved state constitutes
a soft mode that can potentially propagate through the filament's material frame 
around a mexican-hat-type potential.
In the mucus of our airways, which constitutes a soft, 
porous 3D network, glycan chains are omnipresent
and influenza's spike proteins  are known to  efficiently bind and cut them. 
%We show that when coupled to such a non-equilibrium chemical reaction,  
%the curved toro-elastic state can attain a spontaneous rotation 
We next show that such a non-equilibrium enzymatic reaction can 
%for sufficient enzymatic activity, 
induce spontaneous rotation of the curved state, leading to a whole body reshaping 
propulsion similar to -- but different from -- eukaryotic flagella and spirochetes.
} \\
\end{tabular}

 \end{@twocolumnfalse} \vspace{0.6cm}

  ]
%%%END OF TITLE, AUTHORS AND ABSTRACT%%%

%%%FONT SETUP - please do not change any commands within this section
\renewcommand*\rmdefault{bch}\normalfont\upshape
\rmfamily
\section*{}
\vspace{-1cm}

%%%FOOTNOTES%%%

\footnotetext{\textit{$^{a}$~Institute for Theoretical Physics, Heidelberg University, Philosophenweg 19, 69120 Heidelberg, Germany; E-mail: f.ziebert@thphys.uni-heidelberg.de}}
%Address, Address, Town, Country. Fax: XX XXXX XXXX; Tel: XX XXXX XXXX; E-mail: xxxx@aaa.bbb.ccc

\footnotetext{\textit{$^{b}$~BioQuant, Heidelberg University, Im Neuenheimer Feld 267, 69120 Heidelberg, Germany}}
%Address, Address, Town, Country. 

\footnotetext{\textit{$^{c}$~Department of Chemical Engineering and Biotechnology,
University of Cambridge, Cambridge CB3 0AS, UK}}
%Address, Address, Town, Country. 

\footnotetext{\textit{$^{d}$~Institut Charles Sadron UPR22-CNRS, 67034 Strasbourg, France}}
%Address, Address, Town, Country. 

\footnotetext{\textit{$^{e}$~Institute Theory of Polymers, Leibniz-Institute of Polymer Research, D-01069 Dresden, Germany; E-mail: kulic@unistra.fr}}
%Address, Address, Town, Country. 

%%%END OF FOOTNOTES%%%

%%%MAIN TEXT%%%%
\section{Introduction}

\begin{figure*}[t]
	\centering
	\includegraphics[width=.9\textwidth]{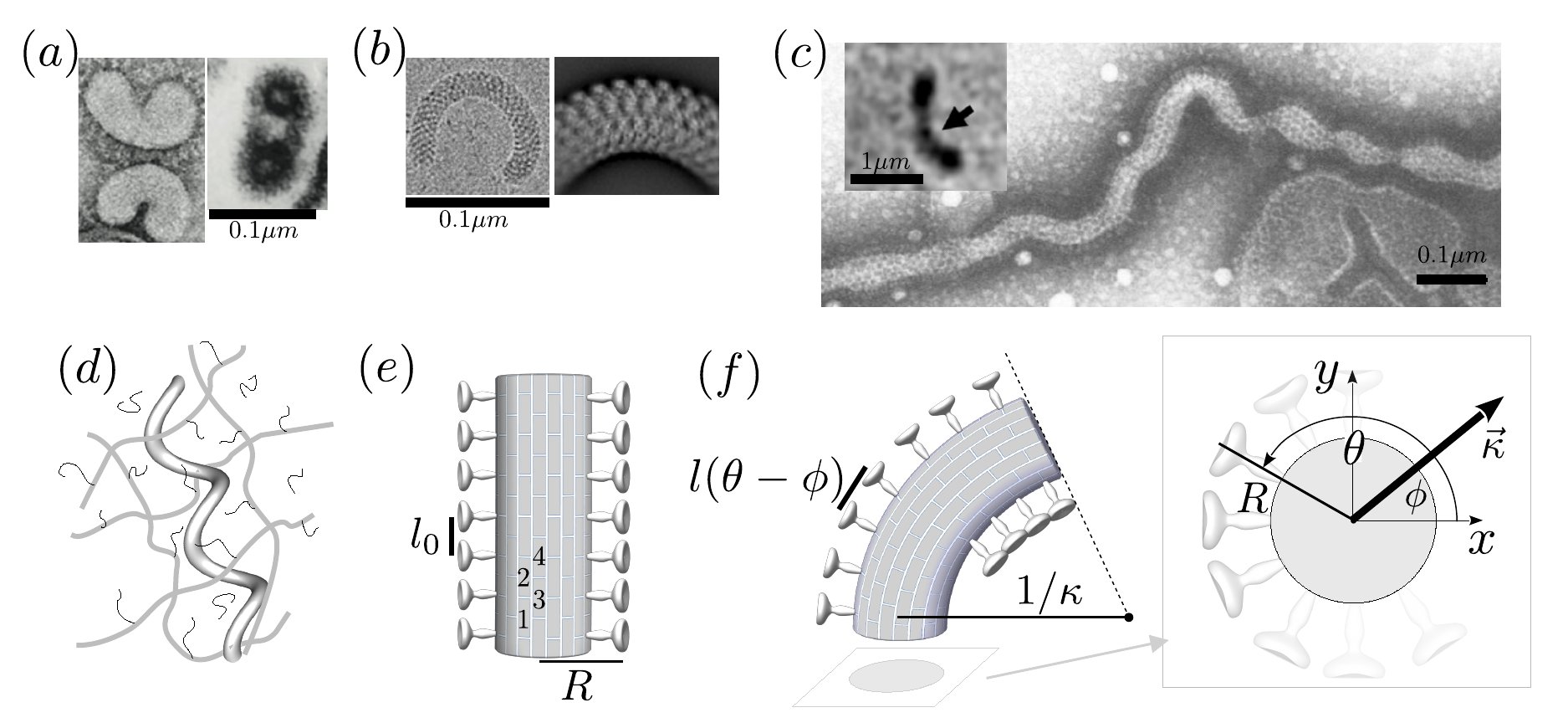}%{Fig1+2.pdf}
	\caption{	 \label{fig1_2} 
	     Emergent toroidal shapes of various biofilaments (a-c), 
	     a filamentous influenza virus in its natural environment -- the mucus  (d), 
	     and geometry of the virus and the toroidal mode (e, f). 
		a) Torovirus \cite{Torovirus-Pic}.
		b) The bacterial flagellar hook \cite{Bact-Hook}.
		c) Influenza C: long filamentous viruses 
		display both straight and oscillatory shapes \cite{IVC-Bending}; 
		the inset shows a shorter, curved Influenza C that is actively rolling on a 
		glycan-coated surface \cite{Sakai_Saito_IVC}.
		(d) The mucus environment that the influenza virus has to bypass forms 
		a loose protein fiber network decorated with sticky glycan chains. 
		(e) In the considered filamentous virus, spike proteins are arranged on a 
		typically staggered column lattice with spacing $\l_0$ 
		and effective radius $R$ (including the spikes). 
		(f) Curving the lattice by a curvature $\kappa$ allows some spikes to interact
		at the inside of the bend (left).
		Which particular spikes can interact depends on their 
		orientation angle $\theta$ (right) with respect to the  
		angle $\phi$ of the 
		curvature vector $\vec{\kappa}$.  
	}
\end{figure*}

Thin fibers are common structural elements in biology,
from biofilaments to body shapes of viruses and bacteria.
Notably, biological filaments 
like the bacterial flagellar hook \cite{Berg-Flagella,Bact-Hook,Namba-2}, 
microtubules \cite{MT1,MT2}, and intermediate filaments \cite{IF-Coiling} 
exhibit an unusual, common motif: their straight ground state can become unstable,
reshaping them into polymorphic toroids and superhelices. 
Not less surprisingly, when driven out of equilibrium, 
stimuli-responsive but otherwise straight fibers 
can acquire a dissipative toroidal steady state
of spontaneous rotation \cite{Baumann,Bazir}. 
Visually inspecting the shapes of filamentous viruses 
like Influenza C (IV-C) \cite{Badham_Rossman,IVC-Bending} or Torovirus \cite{Torovirus-Review},
see Fig.~\ref{fig1_2}a), c), the natural question arises if viral envelopes 
posses a similar symmetry-broken, toroidal ground state.
And as viruses like influenza A (IV-A) and C 
\cite{Sakai_Saito_IVA,Sakai_Saito_IVC,virusPRL}, 
as well as paramyxovirus \cite{paramyxovirus}
have been shown to be able to roll on surfaces due to
their spike proteins' catalytic activity,  
can their toroidal state be driven catalytically and used for virus propulsion? 
 
In the following we consider a plausible model for the emergence of 
curved states, that arise via a spontaneous symmetry breaking, 
represent a soft mode around the virus axis and which 
we call the ``toro-elastic'' state of the viral envelope. 
It relies on surface switchability %\str{at the interfaces}
via spike protein contacts
and a mismatch-stress with layers further inside the virus. 
Notably, recently resolved surface structures of influenza C \cite{IVC-Spike-Lattice} 
have shown extensive spike-spike contacts, 
establishing a hexagonal lattice on the envelope. 
The switch corresponds to the making-and-breaking of any reversible physical bond
-- including hydrogen bonds, screened-electrostatic, van der Waals bonds -- 
as well as protein-tail-mediated interactions. 

After demonstrating the mechanism giving rise to the toro-elastic state, 
we explain how its continuous ground state 
%similar to a built-in 
can act as a ``wheel within the material" \cite{Baumann}, 
mirroring the 
``universal joint" model of the bacterial flagellar hook \cite{Berg-Flagella,Namba-2,Bact-Hook}.

Finally, we consider such a filamentous virus with a toroidal mode
embedded into a loose 3D mucus network, cf.~Fig.~\ref{fig1_2}d).
In this environment, polymeric ligands 
that can be bound to and cut by the enzymatic spikes on the viral envelope
can induce dynamic force imbalances.
This situation reflects what influenza is confronted within the mucus 
of our airways, a highly viscous environment containing polymers 
with sugary ends (glycans) the virus spikes interact with. 
We show that 
the cutting of glycans by enzymatic spikes can force the 
toroidally curved state to spontaneously rotate, 
once enzymatic activity surpasses a certain threshold.   
This 
dynamic shape rotation should allow the virus to move 
at small but noticeable speeds, and could explain why 
catalytic activity increases mucus penetration so strongly for influenza
\cite{Cohen,NA_through_mucus2}.

\section{Toro-elastic state}
\label{sec:toro}
 
We focus here on influenza C, whose spike protein is called
Hemagglutinin-esterase-fusion protein (HEF), for which spike-spike interactions
have been experimentally evidenced \cite{IVC-Spike-Lattice}. We
consider a virus whose surface is covered with radial spikes 
%-- in case of influenza C called Hemagglutinin-esterase-fusion protein (HEF) --
with  initial axial spacing $l_{0}$
(see Fig.~\ref{fig1_2}d). If the virus curves, 
variations of strains $\varepsilon$ and thus distances along the axis
between spikes are amplified proportional
to the effective radius of the cylindrical virus, $R$, 
and the centerline curvature $\kappa$ like 
$l(\kappa)\approx\varepsilon l_{0}=\kappa Rl_{0}$. 
In addition, there is a geometric ``shortcut factor'' for finite deformations:
For spiky surface elements that interact along the shortest spatial distance,
this distance is given by the secant line, see appendix A for details. 

Introducing the vectorial curvature perpendicular to the 
axis $\vec{\kappa}=(\kappa_{x},\kappa_{y})=\kappa(\cos\phi,\sin\phi)$
with orientation angle $\phi$ and magnitude $\kappa$, 
and parameterizing the material-fixed spike positions with angle $\theta$
(with $\theta=\phi$ corresponding
to the orientation of the curvature vector, i.e.~towards the bend),
the spacing between surface contacts along the axis 
is given by, see Fig.~\ref{fig1_2}f and appendix A,
\begin{equation}
l(\kappa,\theta-\phi)=l_{0}\frac{\sin(\kappa l_{0}/2)}{\kappa l_{0}/2}(1-R\kappa\cos\left(\theta-\phi\right))\,.
\label{eq:l(kappa)}
\end{equation}
Here the first term is the shortcut factor described above 
and the second reflects the dependence on angle differences.
At this point we fix $\vec{\kappa}$ to point in the $x$-direction ($\phi=0$).

\begin{figure*}[t]
	\centering
	\includegraphics[width=.9\textwidth]{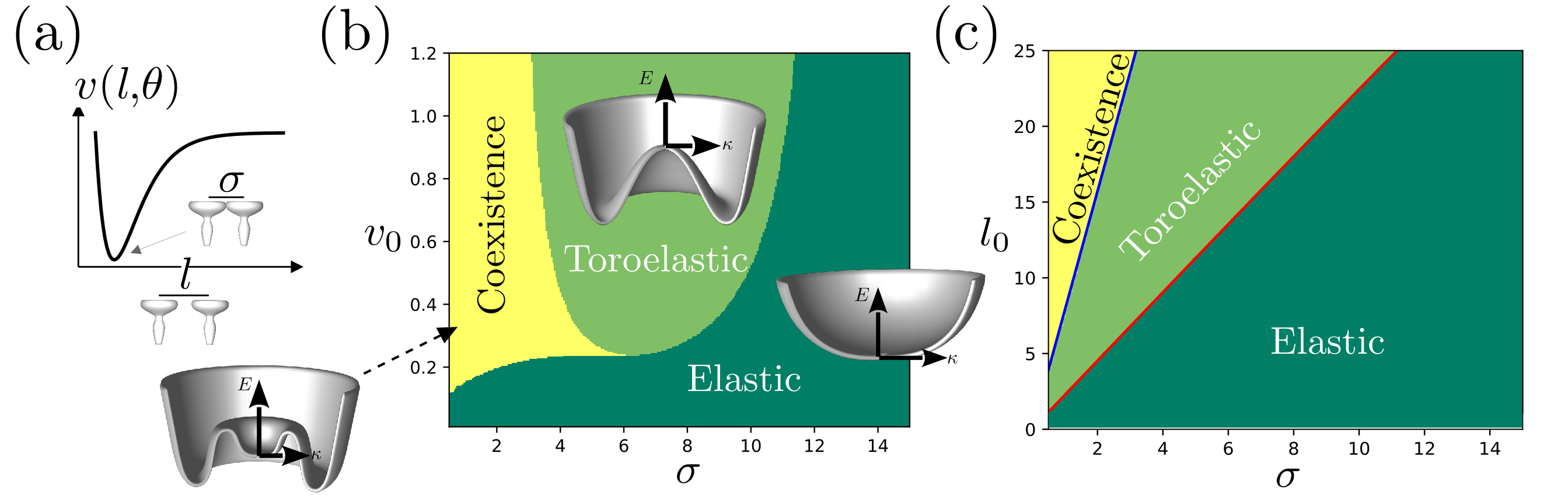}
	\caption{\label{fig3} 
		(a)  Spike interaction potential with a minimum at a finite distance $\sigma$.
		(b)  ``Phase diagram'' in the plane potential depth, $v_0$, 
		vs.~potential minimum position, $\sigma$.
		The straight ground state (dark green) has $B_{eff}>0$ (usual elasticity) 
		and its energy minimum is at $\kappa=0$.
		However, it can become unstable through 
		spike-spike interactions and give rise to a 
		Mexican-hat-type potential (light green), 
		where $B_{eff}<0$ and the energy minimum is at finite $\kappa$ and
		arbitray angle, reflecting a soft mode  (``toro-elastic" mode).
		There also exists a coexistence region (yellow) where $\kappa=0$ 
		and the toro-elastic state are separated by an energy barrier.
		Parameters:  $\sigma$ in nm and  
		$v_{0}$ in $\frac{k_{B}T}{\rm nm}$ for fixed $l_{0}=25$ nm.
		Virus radius $R=50$ nm; bending rigidity $B= 3\,k_BT\,\mu{\rm m}$.
		(c) Phase diagram as function of the unperturbed spike spacing, 
		$l_0$ (in nm) vs.~potential minimum position, $\sigma$ (in nm), 
		for fixed  contact potential strength $v_{0}=1 \frac{k_{B}T}{\rm nm}$,
		other parameters as in b).	
		}
\end{figure*}

The spikes are assumed to be uniformly distributed over the surface, 
cf.~Fig.\ref{fig1_2}f on the right,
and every spike can interact with its axial neighbors
that are within range, 
letting them gain a binding %free 
energy (per unit length), $v(l)$. 
On the flip side, binding along its long axis implies a curving of the cylindrical virus,
which costs elastic bending energy. 
Modeling the latter contribution as a harmonic term with bending stiffness
$B$, the total energy per cross section is hence given by 
\begin{equation}
E=E_{bend}+E_{bind}=\frac{1}{2}B\kappa^{2}
+\int_{0}^{2\pi}v\left(l(\kappa,\theta)\right)d\theta.\label{eq:Energy-Total}
\end{equation}
While HEF-spikes have been shown to strongly interact \cite{IVC-Spike-Lattice},
their detailed binding potential $v(l)$ is yet unknown.
However, as the sum of many small interactions between
different regions along the spikes' surfaces, it will be of short-ranged nature.
Interestingly, 
apart from its magnitude and short ranged 
nature, the detailed form of $v$ turns out to be non-essential 
for the emergence of the toro-elastic state. 
%An instructive potential is the
For demonstration, we here use a linear-exponential potential, 
\begin{equation}
v(l)=-v_{0}\frac{l}{\sigma}\exp\left(1-\frac{l}{\sigma}\right),\label{eq:v_exp-lin}
\end{equation}
as sketched in Fig.~\ref{fig3}a. 
It has a minimum, corresponding to the bound state, at $l=\sigma$ with
$v(\sigma)=-v_{0}$ and quickly flattens for $l>\sigma$. 

Inserting Eqs.~(\ref{eq:l(kappa)}), (\ref{eq:v_exp-lin}) into Eq.~(\ref{eq:Energy-Total}),
the energy can be calculated analytically\footnote{
The result is
\begin{equation}
E=\frac{1}{2}B\kappa^{2}-2\pi v_{0}\frac{a}{\kappa R}e^{\left(1-\frac{a}{\kappa R}\right)}\left(I_{0}(a)-\kappa R\,I_{1}(a)\right),\nonumber
\end{equation}
where $a=\frac{2R}{\sigma}\sin(\kappa l_{0}/2)$
and $I_{0},I_{1}$ are the modified Bessel functions of the first kind.
}.
Importantly, for suitable parameters its shape is 
of Mexican-hat-type, see the inset showing the energy shape in the light green region of Fig.~\ref{fig3}b.
This is what we call the ``toro-elastic'' state: a finite curvature is preferred,
whose direction (i.e~angle $\phi$) is arbitrary and hence corresponds to a soft mode.
Fig.~\ref{fig3}b shows the different energy landscapes/shapes 
when varying the depth, $v_0$, and the minimum position, $\sigma$,
of the potential for fixed spike distance $l_0$ and virus stiffness $B$.
This phase diagram displays: 
(i) an elastic state (dark green) 
where $\vec{\kappa}=0$ is the energy minimum, corresponding to a straight virus; 
(ii) the toro-elastic region (light green), where the minimum is
a circle with $\left|\vec{\kappa}\right|=\kappa_{tor}$
and arbitrary angle $\phi$; 
and (iii) a coexistence region (yellow), where both previous states are local minima,
with a barrier in between.
Fig.~\ref{fig3}c shows a similar phase diagram in the plane spike distance $l_0$ vs.~minimum position
of the potential $\sigma$,
now for fixed potential depth $v_0$.
%to reach the toro-elastic state.

To understand when the straight state %, $\kappa=0$, 
becomes unstable,
we expand the energy up to second order in curvature, yielding 
$E=E_0+\frac{1}{2}B_{eff}\kappa^2$ with 
$E_0=-\frac{2\pi l_{0}v_{0}}{\sigma}e^{1-\frac{l_{0}}{\sigma}}$ and
an effective bending stiffness $B_{eff}=B+B_{bind}$ 
with 
%\begin{equation}
$B_{bind}=2\pi v_{0}l_{0}^{2}e^{1-l_{0}/\sigma}
\left[
\frac{R^{2}}{\sigma^{2}}\left(1-\frac{l_{0}}{2\sigma}\right)+\frac{l_{0}}{12\sigma}\left(1-\frac{l_{0}}{\sigma}\right)
\right]$.  %\,\,\nonumber
%\label{eq:Qudratic-Stiffness}
%\end{equation}
This binding-induced part of the effective stiffness can become negative.
In fact, this can be always satisfied for $l_{0}\gtrapprox2\sigma$
if in addition the  interaction is  sufficiently large,
$v_{0}\gg e^{l_{0}/\sigma}B/l_{0}^{2}$: then $B_{bind}$ overrules $B$
and the filament curves, with higher order terms stabilizing
a finite toroidal curvature $\kappa_{tor}$. 

This toro-elastic state {\it generically} appears 
for short ranged potentials;
%strong surface interaction and sufficiently large (but not too large) spike
%spacing vs.~interaction distance.
a general condition can be determined
by expanding an arbitrary potential $v(l)$ around its straight state $l_0$, see appendix B.
Then for $\frac{R^{2}}{l_{0}^{2}}\gg1$, 
$B_{eff}=B+\pi v''\left(l_{0}\right)l_{0}^{2}R^{2}$, implying that
the curvature
$v''\left(l_{0}\right)$ of the potential must be sufficiently negative,
\begin{equation} 
v''\left(l_{0}\right)<-\frac{B}{\pi l_{0}^{2}R^{2}}.
\end{equation} 
This is in fact fulfilled by many short ranged potentials, 
when crossing over from the minimum to the plateau at large distances.

The virus curvature in the toro-elastic state adjusts such that a significant
portion of the cross-section (typically half of it), benefits from
the surface interactions. For that, the elements of the surface need
to shift from their unperturbed distance $l_{0}$ to $\sigma$, 
i.e.~by an amount $l_{0}-\sigma\sim\kappa_{tor}Rl_{0}$, which 
yields an estimate for the curvature of the toro-elastic state 
\begin{equation}
\kappa_{tor}\sim R^{-1}\left(1-\sigma/l_{0}\right).
\label{kap_tor_est}
\end{equation}
The stiffness of the toroelastic state (i.e.~the curvature
orthogonal to the rim of the ``Mexican hat'') %\str{, cf.~Fig.~\ref{fig3}b}  ), 
is dominated by surface interactions. 
To estimate it, we assume a curvature variation $\delta\kappa$
around $\kappa_{tor}$, such that the binding distance changes by
$\sim\sigma$, 
implying $\delta\kappa\sim\frac{\sigma}{Rl_{0}}$, 
and the binding energy by $\delta E\approx |E_0|$.
Equating
$\frac{1}{2}K_{tor}\left(\delta\kappa\right)^{2}\approx
|E_0|$
then yields 
%\str{allows to estimate the stiffness of the toroelastic state to be}
\begin{equation}\label{K_tor_est}
K_{tor}\approx4\pi\left(\frac{l_{0}}{\sigma}\right)^{3}e^{1-\frac{l_{0}}{\sigma}}R^{2}v_{0}
\end{equation}
up to subdominant contributions from bending elasticity. %neglected here. 

Considering numbers, typical scales
are $l_{0}=25$ nm, $\sigma=10$ nm and %virus radius
$R=50$ nm; 
%a reasonable estimate for the contact 
potential strength around $v_{0}=1 \frac{k_{B}T}{\rm nm}$.
The bending rigidity of a virus is hard to measure \cite{Veigel},
but considering the virus as a membrane tube we estimate 
$B=2\pi R\cdot10 k_B T\simeq 3\,k_BT\,\mu{\rm m}$.
Using these values, one finds
$B_{bind}\simeq-5.7\,k_BT\,\mu{\rm m}$, hence indeed $B_{eff}<0$ is at reach.
Further we estimate 
from Eqs.~(\ref{kap_tor_est}), (\ref{K_tor_est}) that
$\kappa_{tor}\simeq(100\,{\rm nm})^{-1}$  and 
the  toro-elastic stiffness
$K_{tor}\simeq100\,k_BT\,\mu{\rm m}$,
which is much larger than the usual bending rigidity.

\section{Toroelastic state driven by ligand dynamics} 

Having seen how the toro-elastic, curved state arises,
we now show how the orientation angle of the curved state 
-- representing a soft mode due to the Mexican-hat-type potential --
can be driven to rotate in an out-of-equilibrium situation.
For this we assume that the virus is embedded in a solvent that contains
a substantial concentration of glycan-coated polymers 
which the virus spikes' can attach to and enzymatically cut.
This reflects the situation in the mucus, which is a gel/network of
long and heavily (up to 80\%) glycosilated mucins \cite{mucus_review},
having a pore size of 100-500 nm \cite{mucus_pore_size}, 
hence larger than the virus diameter.    
For simplicity, we model this situation via di-glycan "dumbbells" floating in solution,
representing attachment to glycans from the same mucin backbone.
We would like to stress that we consider the virus to be embedded
in a 3D environment,  cf.~Fig.~\ref{fig1_2}d). This should be contrasted
to the catalytic spike-induced motion of viruses on surfaces
\cite{Sakai_Saito_IVA,Sakai_Saito_IVC,Fletcher}. 
In the latter case, strategies like the burnt-bridge mechanism,
employing the trail of cut glycans behind to rectify the motion,
or rolling \cite{virusPRL, virusPRE} can be used.
In contrast, in the absence of surfaces both mechanisms are not possible
or not effective (with the trail in 3D having a mathematical measure close to zero).

\begin{figure}[t]
	\centering
	\includegraphics[width=.99\linewidth]{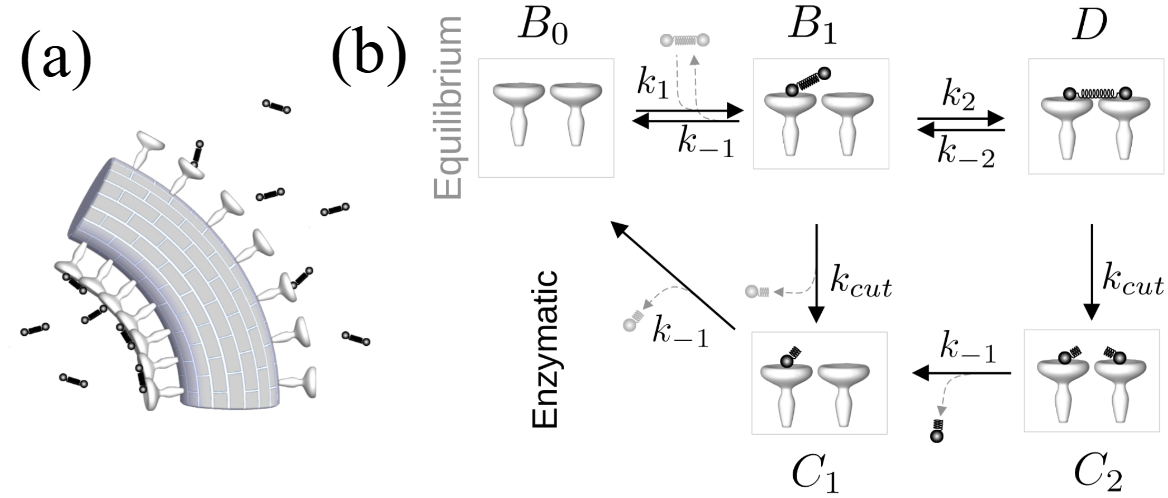}
	\caption{\label{fig4} 
		(a) Interaction of tension-inducing polymeric ligand springs (glycan dumbbells) 
		with the virus spikes. 
		(b) The reaction kinetics of equilibrium binding and catalytic cutting. }
\end{figure}

\begin{figure}[t]
	\centering
	\includegraphics[width=.48\textwidth]{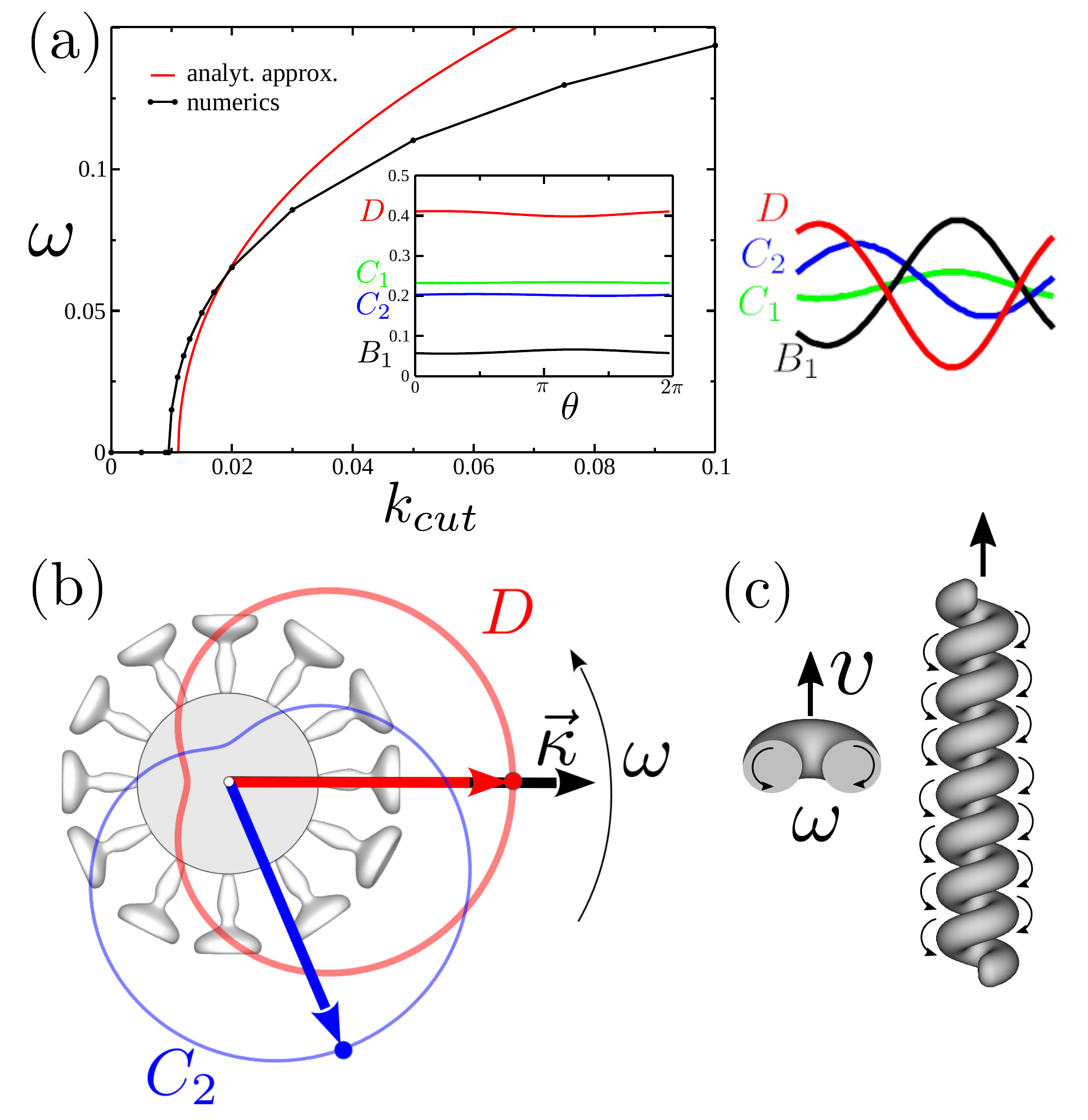}
	\caption{\label{fig5} 
		(a) The toroidal mode's angular frequency 
		as a function of the cutting rate $k_{cut}$ (both in units $\rm{s}^{-1})$.  
		Red curve: analytical result, Eq.~(\ref{eq:omega-condition-1}),
		obtained without friction. Black curve: numerical 
		result with realistic frictional torque coefficient $\xi_h/m_{dr}=0.01$.
		Parameters as described in the main text. Inset: the angular dependence of the steady-state populations.  Right hand side: Scaled and shifted populations for visual comparison. Note their typical phase lag. (b) The polar plot of the double-bound and cut state populations, $D(\theta,t)$ and $C_2(\theta,t)$ respectively, shows their characteristic phase-shift determining the direction of toroidal rotation. Polar plots are schematic and not to scale. (c) The latter kinetics induces a uniform, shape-invariant rotation of a short toroidal virus 
		or an elongated helical virus with frequency $\omega$ . When embedded in a viscous fluid the rotation leads to a 
		swimming kinematics similar to fluid vortex lines 
		\cite{Pozrikidis,Thaokar}.
	}
\end{figure}

%{\color{red} (FZ: I tried to give a simple argument already here:)}  
The argument for rotation involves two steps:
first, when glycan dumbbells attach to two axially neighboring spikes,
they  introduce longitudinal tensions on top of the toro-elastic state, creating a torque. 
Second, the cutting of the glycan dumbbells leads to a release of tension
with a certain dead time for new binding, since cut glycans have yet to detach.
Like that, torque is maintained by breaking of detailed balance 
and the dead time gives the system a ``directional memory'', 
once a direction of rotation is picked  \footnote{ Jumping bit ahead , as we will see in the following section, the direction of rotation is set by the phase-shift of certain ligand-binding populations introduced in Fig. \ref{fig4} and plotted in Fig.\ref{fig5}}.

To calculate the torque due to dumbbell attachment,
we assume the di-glycans to be harmonic
polymer springs with spring constant $S$ and %a 
vanishing preferred length. 
They can be unbound (state $B_0$), bound to one spike (state $B_1$)
or to two neighboring spikes -- along the backbone -- 
defining the double bound state $D$,
see the upper part of the reaction scheme in Fig.~\ref{fig4}b.
In the double bound state, 
glycans stretch out to a non-zero length given by Eq.~(\ref{eq:l(kappa)}),
with $\phi$ the orientation of the curvature %vector 
$\vec{\kappa}$
and $\theta$ the angular position of the spike it attaches to
in the considered cross-section.

Note that, for simplicity, we consider here only \textit{relevant} spike-glycan interactions, 
which are the ones along the virus axis, or having, when projected, a component along the axis.
Couplings along the azimuthal direction are neglected, since they do not contribute 
to axial bending. Note that this simplification introduces effective, phenomenological binding rates.

As the spikes are assumed to be uniformly distributed over the surface,
all states have to be described by a function of the azimuthal angle $\theta$.
In general, a cross-section will have an azimuthal imbalance 
of the (double-)bound density $D\left(\theta\right)$. 
The total glycan stretching energy per cross-section %for a virus segment 
of thickness $l_{0}$ then reads
%\begin{equation}
$E_{str}\left(D\right)  =\frac{S}{2}\int D\left(\theta\right)l^{2}(\kappa,\theta)$.
%\end{equation}
Assuming small curvatures $\kappa R,\,\kappa l_{0}\ll1$
and applying a mode ansatz $D(\theta)=D_0+D_s\sin\theta+D_c\cos\theta$,
to linear order 
$E_{str}\left(D\right) \approx \frac{S}{2}l_{0}^{2}\left(D_{0}-R\kappa\left(D_{c}\cos\phi+D_{s}\sin\phi\right)\right)$.
The corresponding torque (per $l_{0}$ section) 
is given by $M_{dr}=-\frac{\partial E_{str}}{\partial\phi}$  
and in the frame co-moving with the toro-elastic curvature 
evaluates to
\begin{equation}
M_{dr}=\frac{S l_{0}^{2}}{2}R\kappa D_{s}=m_{dr}D_s.\label{eq:Torque}
\end{equation}
This is the driving torque from the inhomogeneously distributed double bound glycans,
with $m_{dr}$ the characteristic torque scale.
Assuming a Kuhn length of a glycan chain of $b\sim0.5$ nm \cite{Li_glycanKuhn}
and a number of bonds of $N\sim40$ or larger one expects typical
spring constants $S\simeq\frac{k_{B}T}{bN^2}\sim0.01-0.1\frac{k_{B}T}{{\rm nm}^{2}}$. Together with
the above parameters and $\kappa=\kappa_{tor}$,
we estimate $m_{dr}\simeq2-20\,k_{B}T$, which is much larger than 
the hydrodynamic resistance of such a small cross-section\footnote{
Note that this value is large, considering that it acts on a short
section of size $l_{0}$. As a reference, the rotational friction at an angular
frequency $\omega=2\pi {\rm s}^{-1}$ in a rather viscous medium 
of $\eta=0.1\,{\rm Pa\,s}$ (100 times of water)
for $R=50\,{\rm nm}$ is approximately $M\approx0.05 k_{B}T$,
i.e.~largely subdominant compared to the linker-induced torque $M_{dr}$
for not too low $D_s$.
}. In view of this, we will first study the case, where resistance can be completely 
neglected, implying that the total torque is zero. It should be however noted that the 
attachment-detachment dynamics to the sticky mucin network elements of the mucus 
can contribute an additional friction term due to dragging a portion of the mucins 
along with the virus. Therefore, in a next step, we will add the (rotational) resistance perturbatively.

To describe the possible steady-state torque generation,
we now consider the enzymatic glycan cutting activity of the HEF spike.
The %\str{initial equilibrium} 
reaction scheme just discussed
has to be extended by two more states associated to the cutting reaction, 
as sketched in the lower part of Fig.~\ref{fig4}b):
First, a double bound glycan can be cut into two, 
to a state $C_2$ that still blocks both spikes for further attachments. 
If one of these cut parts detaches,
or if a single bound glycan is cut, this yields state $C_1$.
Importantly, only if all cut parts have left, one recovers state $B_0$ 
and the spike can be ``reused'' for attachment. 

The full reaction scheme  of Fig.~\ref{fig4}b translates into
\begin{eqnarray}
%\frac{d}{dt}B_{0} & =-k_{1}GB_{0}+k_{-1}B_{1}+k_{-1}C_{1}\\
\frac{d}{dt}B_{1} & =&k_{1}GB_{0}-\left(k_{-1}+k_{2}\left(\theta-\phi\right)
+k_{cut}\right)B_{1}+k_{-2}D,\nonumber\\
\frac{d}{dt}D & =&k_{2}\left(\theta-\phi\right)B_{1}-k_{-2}D-k_{cut}D,\nonumber\\
\frac{d}{dt}C_{1} & =&k_{cut}B_{1}+k_{-1}C_{2}-k_{-1}C_{1},\nonumber\\
\frac{d}{dt}C_{2} & =&k_{cut}D-k_{-1}C_{2},
\label{DynamicEqns}
\end{eqnarray}
where all considered states are functions of $\theta$.
An additional equation for the unbound state $B_0$ can be eliminated 
via $B_{0}+B_{1}+C_{1}+C_{2}+D=1$, since all quantities
are site population fractions.
%{\color{red} (FZ: explained in words:)}
These dynamic equations can be directly deduced from the scheme in  Fig.~\ref{fig4}a.
Exemplarily, the equation for the single bound state $B_1$ has a gain term from 
the attachment of a glycan ($G$) to a free spike ($B_0$) with rate $k_1$;
loss terms proportional to $B_1$ and reflecting detachment (with rate $k_{-1}$), 
attachment to a second spike (with rate $k_2$) and getting cut (with rate $k_{cut}$),
and another gain term if a double bound state ($D$) detaches from the adjacent spike
(rate $k_{-2}$).
Importantly, the double binding rate $k_{2}\left(\theta-\phi\right)$
depends on the position relative to the curvature vector:
for a bent virus, di-glycans bind more easily at the inside
of bent regions where they stretch less, cf.~Fig.~\ref{fig4}a).

We now consider the steady-state, where the virus curvature $\vec{\kappa}$
is rotating with constant angular frequency $\omega$
-- that has to be determined, and could of course be zero -- 
implying $\phi(t)=\omega t$,
and where $\kappa=const.$ as suggested by the high stiffness of the toroidal state estimated above.
We hence transform Eqs.~(\ref{DynamicEqns}) into the co-moving frame, $\phi=0$ , where steady states become time independent and the total time derivative transforms into an advective derivative $\frac{d}{dt}\rightarrow\omega\frac{\partial}{\partial\theta}$.

The double binding rate, $k_{2}\left(\theta\right)$ in the co-moving frame,
can be specified as follows:
we expect a relation
\begin{equation}
k_{2}\left(\theta\right)=k_{2}^{0}\,\left(1+\alpha\cos\theta\right),\label{eq:k2}
\end{equation}
with a coupling constant $\alpha>0$ that that will depend on the
stretching constant of the glycan polymer chains that connect neighboring spikes. 
For small curvatures $\kappa R, \kappa l_{0}\ll1$, 
one can relate the binding rates on the two opposing sides 
%\begin{align*}
$	\frac{k_{2}\left(\theta\right)}{k_{2}\left(\theta+\pi\right)}  
	=e^{\frac{S}{2}\frac{-l^{2}\left(\theta\right)+l^{2}\left(\theta+\pi\right)}{k_{B}T}}
	\approx1+\left(\frac{2Sl_{0}^{2}}{k_{B}T}R\kappa\right)\cos\theta$.
%\end{align*}
Matching this result with
Eq.~(\ref{eq:k2}),
$\frac{k_{2}\left(\theta\right)}{k_{2}\left(\theta+\pi\right)}
=\frac{1+\alpha\cos\theta}{1-\alpha\cos\theta}
\simeq1+2\alpha\cos\theta$,
yields the estimate
\begin{equation}
%k_{2}\left(\theta\right)=k_{2}^{0}\left(1+\alpha\cos\theta\right)\,,\,\,
\alpha=\frac{ Sl_{0}^{2}}{k_{B}T}R\kappa
\label{eq:alpha}
\end{equation}
for the dimensionless coupling constant, which in the following
is assumed to be small for simplicity.

The equations for the steady state dynamics in the co-moving coordinate system  
can be solved by applying a
mode ansatz, similar as above for the driving torque.
Now all chemical species 
%can have a constant, a $\cos$- and a $\sin$-mode,
are expressed as
$X=X^{0}+X^{c}\cos\theta+X^{s}\sin\theta$ for
$X=B_1, D, C_1, C_2$.
In the case of negligible fluid friction, and fast binding of the second bond of the dumbbell, 
$k_{2}^{0}\gg k_{others}$,  the solution can be given analytically, see appendix C. 
One finds that $\omega=0$ -- i.e.~no rotation -- is always a solution. 
However, solutions with finite angular velocity, given by
\begin{equation}
\omega^{2}  =\frac{k_{1}Gk_{cut}}{2}\left(1+\sqrt{1+\frac{8k_{-1}^{2}}{k_{1}Gk_{cut}}}\right)-k_{-1}^{2},\label{eq:omega-condition-1}
\end{equation}
bifurcate from the no-rotation branch for sufficiently rapid glycan cutting rate
\begin{equation}
\label{eq:kcut-condition}
k_{cut}>k_{cut}^{c,0}=\frac{k_{-1}^{2}}{3k_{1}G}\,.
\end{equation}
Rephrasing Eq.~(\ref{eq:kcut-condition}) as 
$\frac{k_1 G}{k_{-1}}\frac{k_{cut}}{k_{-1}}>\frac{1}{3}$,
the first ratio demands a sufficiently large effective attachment rate ($k_1G$)
to supply for new $D$ (for large $k_2$, $B_1$ converts to $D$ fast)
and the second a sufficiently large cutting rate to maintain non-equilibrium.
Due to the product, faster attachment can compensate for slower cutting and vice versa.
Notably, the rotating state also requires a sufficiently long ``dead time'' interval 
$t_{dt}=1/k_{-1}$, i.e.~a sufficiently small off-rate $k_{-1}$ compared to $k_{cut}$. 
This is intuitively understandable, as large dead times allow for the maintenance 
of the angular direction (clockwise/anticlockwise rotation), while too short dead times 
would effectively erase the directional memory of the main force-generating 
``power-stroke'' sequence $B_0 \rightarrow D \rightarrow C_{i} \rightarrow B_0$ 
and the time-ordering of the force-generating state $D( \theta,t)$ in the azimuthal 
direction %of the cross-section 
necessary for persistent rotation. 

It is also possible to solve the problem in the presence of the small, hydrodynamic
rotational resistance.
Assuming a hydrodynamic (Stokes) rotational friction
torque $M_h=\xi_h \omega=4\pi\eta R^{2}l_{0}\omega$
and equating it with the driving torque, Eq.~(\ref{eq:Torque}),
one gets
$k_{cut}^{c}=k_{cut}^{c,0}+k_{cut}^{c,1}\left(\eta\right)$. 
The correction is to leading order 
linear in fluid viscosity $\eta$, 
proportional to $\frac{R}{\alpha^2 S l_0\kappa}$
and in addition depends on the chemical rates.
See appendix C for the result, Eq.~(\ref{eq:-10}), and how it is obtained.
 
Fig.~\ref{fig5}a) shows the angular velocity $\omega$
as a function of the cutting rate $k_{cut}$: the red curve shows
(the positive branch of) Eq.~(\ref{eq:omega-condition-1}) and
the black curve are numerical results including the frictional torque.
Since the kinetic parameters of IV-C have not been characterized, 
we had to resort to estimates for IV-A here:
$G=3\,{\rm mM}$; $k_1=0.1{\rm mM}^{-1}{\rm s}^{-1}$, $k_{-1}=0.1{\rm s}^{-1}$ 
\cite{Sauter1,Sauter2}; 
$k_{2}^0=1{\rm mM}^{-1}{\rm s}^{-1}$, $k_{-2}=0.1{\rm s}^{-1}$. 
Note that cutting rates are estimated %very 
conservatively, IV-A's neuraminidase reaches rates of up to
$k_{cut}=15\,{\rm s}^{-1}$ 
\cite{Adams}.

We have shown that dynamic force imbalances, caused by interactions
of the virus spikes with glycans from the 3D environment,
can lead to a rotation of the toroidal state.
We note that in section \ref{sec:toro}, we considered the toroidal state
to be caused by spike-spike interactions.
Hence the creation of the toroidal state, and its rotation
outside of equilibrium, have different causes: 
spike-spike vs.~spike-glycan interactions.
While the former needs localized, anharmonic interactions,
for the additional imbalance caused by the latter linear spings are sufficient. 
Experimental evidence shows strong spike-spike interactions,
at least for IV-C \cite{IVC-Spike-Lattice}.
Nevertheless, one could also imagine the toroidal state to be caused 
(or modulated) 
by spike-glycan interactions as well.
%but then one would have to consider nonlinear springs. 
In that case, the rotation is probably less stable:
when the virus moves, its rear will be exposed to fewer uncut 
glycans, and the helix at the rear may unfold.
Such effects, as well as considering a varying density of the enzymatic spikes 
along the virus backbone -- or a phase-separated one, as in IV-A \cite{Fletcher} --
would be interesting to investigate in the future.

\section{Helix formation}
 
Let us now come back to the helical shapes of IV-C as shown in Fig.~\ref{fig1_2}c,
noting that also filamentous IV-A sometimes displays curvature \cite{Kordyukova}.
So far, we  considered a single cross-section, 
hence the natural question arises how different cross-section are coupled.
As sketched in Fig.~\ref{fig1_2}d as pairs 1-2 and 3-4, 
in general viruses have staggered spike sub-lattices.

Assuming for simplicity two sub-lattices, I and II, these 
experience the mean effective curvature induced by the other sub-lattice over their length scale 
($l$) as  
$\kappa_{II}=\left| \frac{1}{l}\int_{-l/2}^{+l/2}\vec{\kappa}_{I}\left(s\right)ds\right|$ and analogously for $\kappa_I$, implying a smearing-out of sectional curvature over the two lattices. 
For a curvature vector of constant amplitude slowly winding 
with a torsion rate $\tau\ll l^{-1}$ like 
$\vec{\kappa}\left(s\right)=\kappa\left(\cos\left(\tau s\right)\mathbf{e}_{x}+\sin\left(\tau s\right)\mathbf{e}_{y}\right)$, the total energy of the two sub-lattices can be calculated to be, see appendix D,
as 
\begin{equation}
E_{tot}/K_{tor}
=\frac{1}{4}\left(\kappa_{tor}-\kappa\right)^{2}
+\frac{1}{4}\hspace{-1mm}\left[\left(\kappa_{tor}-\kappa\right)+\kappa\frac{l^{2}\,\tau^{2}}{24}\right]^{2}.
\end{equation} 
This energy is positive definite. 
Its ground state is curved and untwisted, 
$\kappa=\kappa_{tor}$ and $\tau=0$. 
The first term is similar to classical bending energy, 
followed by a torsion-bend coupling that is 
quartic for $\kappa=\kappa_{tor}$. 
Notably, for curvatures {\it larger} than optimal, $\kappa>\kappa_{tor}$, 
the torsion becomes bistable with two new preferred states 
$\tau_{1,2}=\pm\frac{\sqrt{24}}{l}\sqrt{1-\kappa_{tor}/\kappa}$. 
Hence external bending torques
-- like those due to the above-discussed glycan binding --
will turn the toroidal ground state into a bistable helix.
The symmetry between the two helical states 
is likely broken by a geometrical chirality of the natural spike lattice 
-- an effect neglected here for simplicity.

\section{Discussion and conclusions}
There is an old question from biology asking "why nature does not have wheel and axles". 
In the light of what we know about the emergence of toroidal zero-energy modes 
in the bacterial flagellar hook\cite{Bact-Hook,Namba-2}
(and possibly microtubules\cite{MT1,MT2}), the question can be tentatively answered by: 
Nature has something better -- the "wheel within (the filament)". 
In fact,  to implement 
and maintain bi-component wheel-axle system is hard to achieve biologically,
but a very practical replacement exists: any elastic object equipped 
with a circularly symmetric zero-energy mode. 
Such structures have the great advantage of being single piece continuous structures 
that are containing both an axle (the material frame of the structure), 
and the wheel -- the deformation propagating (rotating) with respect to the material frame 
at virtually no elastic energy cost.  
The simplest of such ZEEMs\cite{Baumann} (zero elastic energy modes) 
seems to be the toro-elastic mode we revisited here for filamentous viruses.

We have shown that such a mode can generically appear for any rough surface spiked cylinder 
-- like a filamentous influenza virus -- for sufficiently strong and short ranged 
spike interactions, 
competing with elastic restoring forces of the elastic core of the cylinder.  
Once this toro-elastic ZEEM emerges and equilibrium detailed balance is somehow broken
-- as by the spike-catalysed glycan-cutting investigated here -- 
it becomes possible for this ``wheel-like'' mode to actively rotate in a particular direction, 
randomly picked by a non-equilibrium bifurcation as investigated here. 
The resulting spatial motion, that could be seen as a shape invariant, 
constant angular velocity ``re-bending" dynamics of the filament perpendicular to its axis 
is similar to the anholonomic body reshaping of a falling cat \cite{Marey, KANEFallingCat} and has been described in various terms and contexts in the literature. 
Other examples include the ``eversion/inversion" for 
self-rolling polymer fibers \cite{Bazir,Eversion-Kai-Li}, 
``wobbling motion" in clamped, superhelical microtubules \cite{MT1} 
and the ``ideal joint" rotation for the bacterial flagellar hook 
driven by the flagellar motor \cite{Namba-2}.   
A related active re-bending/eversion motion in the soft, body reshaping 
bacterium spiroplasma has been been also recently described \cite{Wolgemuth_new}.
	
Furthermore we have shown that the toro-elastic mode in a long filament equipped 
with a sticky surface lattice can induce helix formation in addition to toroids. 
The spontaneous emergence of multi-stable and rearranging helices 
is very common in biological filaments and their origins have been studied 
in various systems including bacterial flagella 
\cite{CALLADINE1976-Bact-Flagellum,Powers,Wada2008Flagella, Stark-Bact-Flagellum}, 
microtubules \cite{MT1,MT2}, intermediate filaments \cite{IF-Coiling} 
and spiroplasma \cite{Wolgemuth_new}. Looking at the wealth of examples, 
one is tempted to suspect that the cylindrically symmetric filament shape 
could be structurally unstable against perturbations and that the superhelical 
filament is the rule rather than the exception in Nature. 
The common theme behind the emergent helicity in these filaments seems to be 
some form of core-shell stress mismatch combined with various forms of 
elastic anharmonicity, that originates from mechanical, geometric or molecular nonlinearities.
In contrast to these previous examples, in the proposed virus model, 
the helical states are still subcritical or "evanescent", 
that is, only upon (arbitrary small) additional stresses, 
the straight state spontaneously breaks the symmetry and picks one 
of two handed helices. However, it appears possible or even likely that 
specific spike-lattice arrangement and chiral spike-spike interactions 
would give rise to more specific, additional effects, so far neglected here. 
This would lead to definite handedness helices in the ground state of the virus 
and be described by similar physics as in the other helical 
systems\cite{CALLADINE1976-Bact-Flagellum,Powers, Friedrich-Bact-Flagellum,Stark-Bact-Flagellum}.

Finally, if -- via the enzymatic drive described above -- the toroidal or helical virus uniformly 
rotates its body (keeping its shape invariant, see Fig.~\ref{fig5}b),
we expect it to self-propel and swim.  
Whether the motion corresponds to a toroidal or a helical rotation 
depends on the presence (or absence) of angular shift between the directions 
of curving of two subsequent cross-sections. 
In the case that all curvature directions are found in a single (osculating) plane, 
an ideal toroidal rotation occurs, see Fig.~\ref{fig5}b) on the left. 
If the plane of curvature is changing along the contour,
the rotation is helical, see Fig.~\ref{fig5}b) on the right.
The swimming kinematics will be that of a helically shaped spinning cable. 
Mathematically this is a vortex line of fixed vorticity $\propto \omega$, 
self-advecting through the fluid \cite{Pozrikidis}.
In leading order the self-propulsion is independent of the (small) helical torsion 
and occurs at typical velocities of spinning tori 
$v\sim\omega\kappa R^{2}$ \cite{RotatingTorus,Thaokar}. 
Note that the predicted shape invariant swimming contrasts body-reshaping swimmers (like spirochetes) 
using propagating bending waves, where the velocity depends on undulation amplitude 
and wavelength but not on radius \cite{Wolgemuth}.

The described motility mechanism should be directly applicable
to IV-C. It could also be relevant for influenza A, where, however, the two 
dynamic properties of the spike 
-- the binding to/unbinding from glycan and the cutting of glycan residues --
are not co-localized in a single HEF spike. In IV-A, these tasks are performed by
two different spike proteins, hemaglutinin (HA) for the binding/unbinding 
and neuraminidase (NA) for the cutting. This makes the surface structure
of IV-A more complex, and in fact, HA and NA are often phase-separated
on the virus surface \cite{Fletcher}, implying that the theory developed here
has to be generalized to inhomogeneous spike distributions.
Our theory is also relevant for the design of synthetic self-propelled objects,
where the adhesion-cutting mechanism has been already implemented, e.g., via DNA-RNA
hybridization (binding/unbinding) and catalytic action of RNase H (selectively cutting
duplexes), to induce surface-based motion \cite{Yehl,Bazrafshan,Kerstin,Korosec}.
The here-described 3D toroidal motion should be at reach for
DNA nano-tubes floating in a solution of RNA, able to bind to two DNAs along the backbone, 
and  RNase H. One could also think of synthetically 
motorizing the flagellar hook \cite{Bact-Hook,Namba-2}.

Coming back to viruses, while IV-C has rarely been studied, 
many studies on IV-A suggest that the enzymatic activity (of NA in case of IV-A)
is important for mucus penetration, both on the 
macroscopic level \cite{NA_through_mucus2,Cohen}, 
as well as by observing filamentous viruses directly \cite{Fletcher}.
For purely spherical viruses, however, it seems that NA is less relevant and it is rather
the mesh size of the mucus that determines the diffusion, as reported in\cite{Kaler}.
This finding is in accordance with our model and the one by Vahey and Fletcher \cite{Fletcher}, 
where only filamentous viruses can profit from enzymatic mechanisms.
We note that the mechanism described in \cite{Fletcher} is ratchet-based,
hence the longer the filament, the slower the motion -- since then the virus 
engages many links, causing increasing friction, proportional to virus length. 
In contrast, the mechanism proposed here would be effective
independent of virus length.

In future studies, the nature of the mucus as a natural habitat environment 
for filamentous influenza viruses needs some deeper considerations. 
Regardless of the detailed motility mechanism, 
the mucus poses a significant barrier to viral motion. 
Depending on its mesoscale morphological features, like its pore size \cite{Kaler} or 
glycan density and type \cite{mucus_review}, 
it will impose different constraints to the viral filament in various shapes and modes of motion. 
Large scale motion through such an environment would not be easily possible 
for a virus that is straight or a piece of a torus, 
while a helical filament could easily “follow its own footprint” 
and reptate its way through the mucus pores.

In conclusion,
filamentous viruses like influenza C, possessing the toro-elastic state
due to axial interactions
can develop helical shapes and 
can attain a spontaneous rotation upon binding and cutting interactions 
with the glycans floating in the mucus. 
The predicted angular velocities are comparable to the rolling angular velocities 
of surface attached viruses (fractions of ${\rm s}^{-1}$ \cite{Sakai_Saito_IVC,virusPRL}).
The expected 
swimming velocity of tens of $\rm{nm}/s$ 
is rather modest compared to swimming speeds of microorganisms,
yet still of the order of one virus size per second
and hence
preferable to staying
immobilized in the extremely viscous, sticky environment faced in the mucus. 
The proposed mechanism adds another possible 
motility mode -- effective in 3D surroundings such as loose mucus networks --
to the existing repertoire of the influenza family and torovirus, 
beyond ratchet-like motion \cite{Fletcher} 
and rolling \cite{Sakai_Saito_IVA,Sakai_Saito_IVC}, which both employ 
glycan-coated surfaces.

\section*{Appendix}

\subsection*{A\quad Derivation of the ``shortcut factor'': surface-surface contact spacing}

\begin{figure}[t!]
\centering
\includegraphics[width=.5\linewidth]{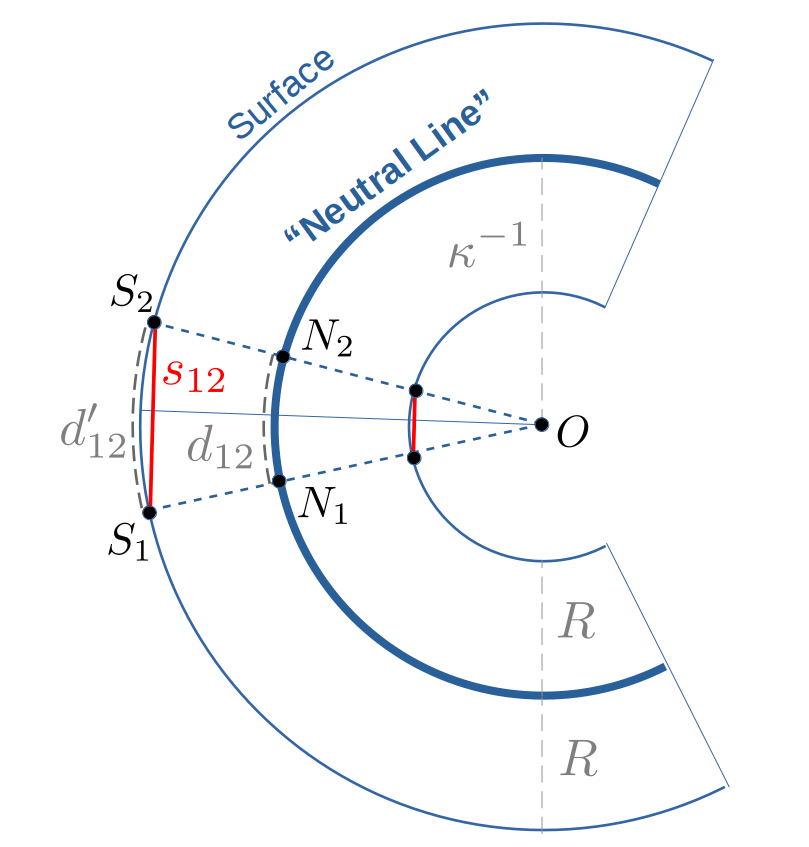}
\caption{\label{shortcut}
Sketch of the geometry of a strongly bent cylinder (the virus capsid/envelope)
and the emergence of the short-cut factor. }
\end{figure}

To derive the formula for the spacing between surface contacts
 from the main text, Eq.~(\ref{eq:l(kappa)}), we
assume that deformations are linear, keeping the
``neutral line'' arc-length contour distances unchanged, 
i.e.~$d_{12}=const.$, independent of $\kappa$. 
After uniformly bending the virus with a center line curvature $\kappa$, 
cf.~Fig.~\ref{shortcut},
the surface elements on the outside
of the bend (found at a radial spacing $R$ from the neutral line)
change their distances along the circular contour proportional to
the curvature $\kappa$ 
\[
d'_{12}=d_{12}\left(1+R\kappa\right)\,.
\]
For spiky surface elements that interact along the shortest spatial
distance, the relevant spatial distance $s_{12}$ between two spikes 
is shorter than the surface arc-length $d'_{12}$: 
it is given by the length of the secant line corresponding 
to an opening angle $\gamma=\frac{1}{2}\angle(S_{1}OS_{2})=\frac{1}{2}\kappa d_{12}$
with length
\begin{align*}
s_{12} =\frac{\sin\left(\gamma\right)}{\gamma}d'_{12}
 =\frac{\sin\left(\frac{1}{2}\kappa d_{12}\right)}{\frac{1}{2}\kappa d_{12}}d_{12}\left(1+R\kappa\right)\,.
\end{align*}
The spatial distance can be simply generalized to any azimuthal orientation
$\theta$ of the bent cylinder (where $\theta=0$ points to the ``inside''
of the bend) such that 
\[
s_{12}\left(\theta\right)=\frac{\sin\left(\frac{1}{2}\kappa d_{12}\right)}{\frac{1}{2}\kappa d_{12}}d_{12}\left(1-R\kappa\cos\theta\right)\,.
\]
With $d_{12}=l_{0}$ and giving the curvature direction
another angle, $\phi$, this is Eq.~(1) from the main text.

\subsection*{B\quad Influence of the potential shape on the emergence of the toro-elastic state}

\noindent\textbf{Gaussian potential.}
We investigated a more general potential of Gaussian form,
\begin{equation}
v(l)=-v_{0}\exp\left( -\frac{(l-\sigma)^2}{2\delta}\right),
\end{equation}
where one can vary the position of the minimum ($\sigma$),
the depth ($v_0$) and in addition the width of the potential well (variance $\delta$)
all independently.

\begin{figure}[t!]
\centering
\includegraphics[width=.95\linewidth]{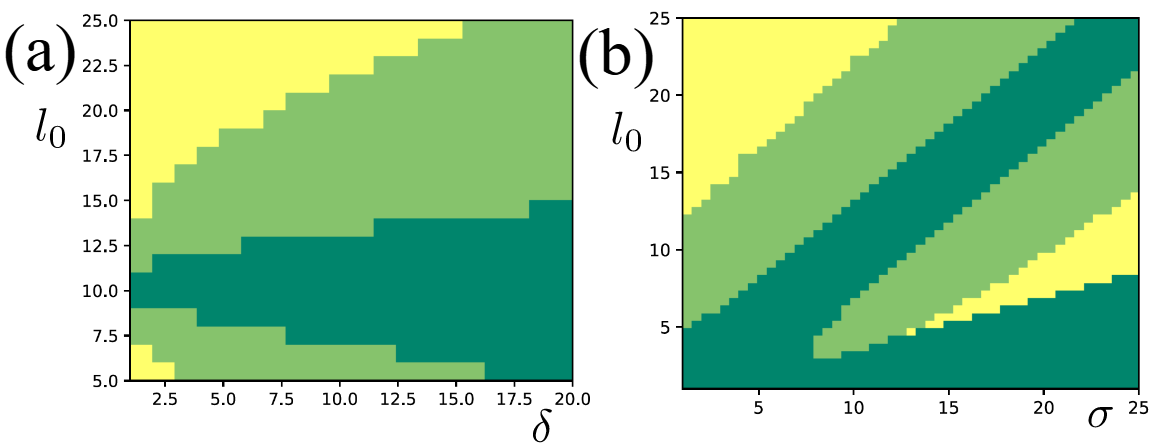}
\caption{Phase diagrams for the Gaussian-type potential. 
a) $l_{0}$ in nm and $\delta$ in nm$^2$ for $\sigma=10$ nm. 
b) $l_{0}$ and $\sigma$ in nm for $\delta=10$ nm$^2$. 
Color-code is according to the energy minima, as in Fig.~\ref{fig3}b),c).
Dark green: elastic state, straight state is stable; 
light green: toro-elastic state, with an energy minimum at finite $\kappa$; 
yellow: coexistence region.
Parameters: virus radius $R=50$ nm; virus bending rigidity 
$B= 3\,k_BT\,\mu{\rm m}$; contact potential strength $v_{0}=1 \frac{k_{B}T}{\rm nm}$.
\label{phasediags_Gauss}
}
\end{figure}

Unfortunately, the energy cannot be calculated analytically anymore.
Fig.~\ref{phasediags_Gauss} shows results 
for the ``phase'' diagrams obtained by numerical minimization. 
The color code is the same as in Fig.~\ref{fig3}.
The left figure shows the plane   $l_0$ vs.~variance $\delta$
for fixed  $\sigma=10$ nm. 
One can see a certain symmetry with respect to  $l_0=10$ nm:
now, the spikes can also induce the toro-elastic state by stretching out
towards the minimum of the potential.
The toro-elastic state is easiest to attain for a short ranged (variance) $\delta$
of the potential, but on the other hand its existence region shrinks for 
small $\delta$ and the region
of coexistence increases.
The right figure shows the plane $l_0$ vs.~$\sigma$ for fixed
 $\delta=10$ nm$^2$. One can see interesting series of transitions.
For instance, choosing $\sigma=15$ nm, upon increasing $l_0$ from a small value
one finds first the elastic state, then coexistence, then the toro-elastic state, 
again an elastic region and a toro-elastic region (and again coexistence
for larger $l_0$, beyond the region visible in the figure).

The additional study of the Gaussian potential shows 
that the shape of the potential of course matters,
but that the three states described in the main text
-- elastic, toro-elastic and coexistence --
are again found. \\

\noindent\textbf{Harmonic Potential.} To shed some light on which
properties the binding potential has to fulfill in order for the toro-elastic instability to occur,
let us study the simplest case of a harmonic binding potential.
Note that a harmonic $v(l)$ is not physically realistic, as every realistic binding potential 
will have a finite range.
Using
%\begin{equation}
$v(l)=a(l-b)^2$
%\end{equation} 
with some spring stiffness $2a$ and preferred length $b$,
the binding energy reads (note that $l=l(\kappa,\theta)$)
\begin{align*}
E_{bind} & =\int_{0}^{2\pi}v(l(\kappa,\theta))d\theta=const+a\int_{0}^{2\pi}l^{2}d\theta
-2ab\int_{0}^{2\pi}l d\theta
\end{align*}
The two integrals can be evaluated
\begin{align}
L_{1}\left(\kappa,\theta\right)&=\int_{0}^{2\pi}l(\kappa,\theta)d\theta=2\pi l_{0}\frac{\sin(\kappa l_{0}/2)}{\kappa l_{0}/2},\nonumber\\
L_{2}\left(\kappa,\theta\right)&=\int_{0}^{2\pi}l^{2}(\kappa,\theta)d\theta=l_{0}^{2}\left(\frac{\sin(\kappa l_{0}/2)}{\kappa l_{0}/2}\right)^{2}\left(2\pi+\pi R^{2}\kappa^{2}\right), \label{L1L2}
\end{align}
and dropping the constant energy shift, one gets
\begin{align*}
E_{bind} 
  =2\pi al_{0}^{2}\frac{\sin(u)}{u}\left(\frac{\sin(u)}{u}\left(1+2\left(\frac{R}{l_{0}}\right)^{2}u^{2}\right)-2\frac{b}{l_{0}}\right)
\end{align*}
with $u=\kappa l_{0}/2$. 
Expanding for small $u$ 
\begin{comment} 
yields
\begin{align}
E_{bind}=2\pi al_{0}^{2}\bigg[&
-\frac{2\,b-l_{0}}{l_{0}}+\frac{\left(l_{0}\,b+6\,R^{2}-l_{0}^{2}\right)\,u^{2}}{3\,l_{0}^{2}}\nonumber\\
&-\frac{\left(3\,l_{0}\,b+120\,R^{2}-8\,l_{0}^{2}\right)\,u^{4}}{180\,l_{0}^{2}}+\cdots
\bigg]
\end{align}
\end{comment}
shows that in the relevant limit $R>l_{0}$ and for any positive preferred length $b$ 
of the harmonic spring, the quadratic term is always positive. 

But in principle, a negative quartic term could still induce an instability.
To investigate this, adding the bending
stiffness term $\frac{B}{2}\kappa^{2}=\frac{B}{2}\left(2\frac{u}{l_{0}}\right)^{2}=\frac{2B}{l_{0}^{2}}u^{2}$ yields the total energy
\begin{equation}
\frac{E_{tot}\left(u\right)}{2\pi al_{0}^{2}}=\gamma u^{2}+\frac{\sin(u)}{u}\left(\frac{\sin(u)}{u}\left(1+2\rho^{2}u^{2}\right)-2\beta\right)
\end{equation}
with the three dimensionless constants 
\begin{align*}
\gamma  =\frac{B}{\pi al_{0}^{4}}\,,\,\,\,
\rho  =\frac{R}{l_{0}}\,,\,\,\,
\beta  =\frac{b}{l_{0}}
\end{align*}
where $\rho>1$, typically even $\rho\gg1$.

Taylor expanding for small $u$ yields, neglecting a constant,
\begin{align*}
\frac{E_{tot}\left(u\right)}{2\pi al_{0}^{2}}=%1-2\,\beta+
\frac{\left(3\,\gamma+6\,\rho^{2}+\beta-1\right)\,u^{2}}{3}-\frac{\left(120\,\rho^{2}+3\,\beta-8\right)\,u^{4}}{180}.
%+ & O\left(u^{6}\right)\,.
\end{align*}
When the (negative) quartic and the (positive) quadratic term become comparable,
we expect a barrier $u\approx u_{bar}$ to occur which
for the special case of $\beta=0$ can be given as
\begin{align*}
u_{bar} \simeq\left(\frac{\,3\,\gamma+6\,\rho^{2}-1}{2\,\rho^{2}-\frac{2}{15}}\right)^{1/2}
%  >_{\text{(typically)}}\left(\frac{\,3\,\gamma+6\,\rho^{2}}{2\,\rho^{2}}\right)^{1/2}\gg1.
\end{align*}
which typically is $\gg1$.
That means in practice this case is not relevant, as excluded volume effects
should enter way before, at $u$ of order 1.

In summary, we have just shown that for a harmonic potential, the straight state increases in
stability and cannot show an instability. \\

\noindent\textbf{Criterion for toro-elasticity for a general potential.}
Let us now assume a general potential $v\left(l\right)$ 
and investigate, when the straight state can become unstable. 
For that we expand $v\left(l\right)$ around its straight state $l=l_{0}$,
\begin{equation}\label{vexpansion}
v(l)=v_{0}+v'\left(l_{0}\right)\left(l-l_{0}\right)+\frac{1}{2}v''\left(l_{0}\right)\left(l-l_{0}\right)^{2}+...
\end{equation}

The binding energy can  again be evaluated,
expanding the functions $L_1, L_2$ defined in Eq.~(\ref{L1L2})
in powers of $\kappa$.
\begin{comment}
\begin{align*}
E_{bind}  &=\int_{0}^{2\pi}v(l(\kappa,\theta))d\theta \\
  &=2\pi v_{0}-2\pi l_{0}v'\left(l_{0}\right)+\frac{2\pi}{2}l_{0}^{2}v''\left(l_{0}\right)
  +\left(v'\left(l_{0}\right)-l_{0}v''\left(l_{0}\right)\right)L_{1}\left(\kappa,\theta\right)
  +\frac{1}{2}v''\left(l_{0}\right)L_{2}\left(\kappa,\theta\right)
\end{align*}
with the integrals
\begin{align*}
L_{1}\left(\kappa,\theta\right)=\int_{0}^{2\pi}l(\kappa,\theta)d\theta % & =2\pi l_{0}\frac{\sin(\kappa l_{0}/2)}{\kappa l_{0}/2}
&\approx2\,l_{0}\,\pi-\frac{l_{0}^{3}\,\pi\,\kappa^{2}}{12}+\cdots\\
L_{2}\left(\kappa,\theta\right)=\int_{0}^{2\pi}l^{2}(\kappa,\theta)d\theta %& =2\pi l_{0}^{2}\left(\frac{\sin(\kappa l_{0}/2)}{\kappa l_{0}/2}\right)^{2}\left(1+\frac{1}{2}R^{2}\kappa^{2}\right)
&\approx2\,l_{0}^{2}\,\pi+l_{0}^{4}\,\pi\left(\frac{R^{2}}{l_{0}^{2}}-\frac{1}{6}\right)\,\kappa^{2}+\cdots
\end{align*}
\end{comment}
Again dropping constant energy contributions, we get 
\begin{equation}\label{generic_crit}
E_{bind}=\frac{1}{2}l_{0}^{4}\,\pi\left[-\frac{v'\left(l_{0}\right)}{6l_{0}}+v''\left(l_{0}\right)\frac{R^{2}}{l_{0}^{2}}\right]\,\kappa^{2}+...
\end{equation}
Interestingly, going on in the expansion of Eq.~(\ref{vexpansion}),
the contribution from $v'''\left(l_{0}\right)$ cancels in $O(\kappa^2)$.
Furtheron, it is easy to check that, using Eq.~(\ref{generic_crit}) for
the linear-exponential potential, Eq.~(\ref{eq:v_exp-lin}), exactly
recovers the criterion discussed in the main text.
This both corroborates that the contributions from
$v'\left(l_{0}\right)$ and $v''\left(l_{0}\right)$
are the most relevant contributions.

For the toro-elastic state to emerge, 
the prefactor of the term $\propto\kappa^2$ in Eq.~(\ref{generic_crit}) 
has to be negative. In general, the second term is the dominant one, 
due to its large geometric prefactor $\frac{R^{2}}{l_{0}^{2}}\gg1$.
As a generic neccessary condition for the toro-elastic instability
-- in the relevant parameter range for a thin virus -- 
one can hence state that 
$v''\left(l_{0}\right)<0$
should hold.
A harmonic potential is strictly convex and $v''\left(l_{0}\right)>0$
forbids the instability, as shown before.
But $v''\left(l_{0}\right)<0$ is of course possible for a short ranged 
potential in its \textit{concave region}. 
In fact, all short ranged potentials have a concave region,
namely when crossing over from the minimum to the zero-level plateau
at large distances. 

Including the bending rigidity,
the total potential reads
\[
E_{tot}=\frac{1}{2}\left(B+\,\pi v''\left(l_{0}\right)l_{0}^{2}R^{2}-\frac{\pi}{6}l_{0}^{3}\,v'\left(l_{0}\right)\right)\kappa^{2}+...
\]
and  for $\frac{R^{2}}{l_{0}^{2}}\gg1$  is dominated by the first two terms. 
The toroelastic state then emerges for 
\begin{equation}\label{toro-criterion}
v''\left(l_{0}\right)<-\frac{B}{\pi l_{0}^{2}R^{2}},
\end{equation}
as stated in the main text.

\subsection*{C\quad Perturbation theory for the rotating state}

We now consider the case where the toro-elastic curvature $\vec{\kappa}$ 
may rotate in the lab frame with angular velocity $\omega$,
due to the chemical reactions (attachment and cutting of glycan dumbbells).
We place ourselves in the co-moving coordinate system where $\vec{\kappa}$
is stationary and can rewrite the dynamics by replacing the time derivative in terms of the advective derivative
$\frac{d}{dt}\rightarrow\frac{\partial}{\partial t}+\omega\frac{\partial}{\partial\theta}$.
As we are only interested in the steady state, the partial time derivative
drops out and we are left with the replacement rule $\frac{d}{dt}\rightarrow\omega\frac{\partial}{\partial\theta}$ for the chemical
reactions in the main manuscript.
Eliminating $B_{0}$ by using $B_{0}=1-B_{1}-C_{1}-C_{2}-D$,
Eqs.~(\ref{DynamicEqns}) in the stationary, co-moving frame now read
\begin{align}
\omega\frac{\partial}{\partial\theta}B_{1} 
%&=&+k_{1}G\left(1-B_{1}-C_{1}-C_{2}-D\right)-\left(k_{-1}+k_{2}\left(\alpha,\theta\right)+k_{cut}\right)B_{1}+Dk_{-2}\nonumber\\
 =&+k_{1}G-k_{1}G(C_{1}+C_{2})-\left(k_{1}G+k_{-1}+k_{2}\left(\alpha,\theta\right)+k_{cut}\right)B_{1}\nonumber\\
 &+(k_{-2}-k_{1}G)D\,,\nonumber\\
\omega\frac{\partial}{\partial\theta}D 
=&+k_{2}\left(\alpha,\theta\right)B_{1}-k_{-2}D-k_{cut}D\,,\nonumber\\
\omega\frac{\partial}{\partial\theta}C_{1} 
=&+k_{cut}B_{1}+k_{-1}C_{2}-k_{-1}C_{1}\,,\nonumber\\
\omega\frac{\partial}{\partial\theta}C_{2}
=&+k_{cut}D-k_{-1}C_{2}\,.
\end{align}

Now we make the simple mode ansatz already mentioned in the main part: 
all chemical species $X=B_1, D, C_1, C_2$
can have a constant, a cos- and a sin-mode,
\begin{equation}
X=X^{0}+X^{c}\cos\theta+X^{s}\sin\theta.
 \end{equation}
The amplitudes $X^0, X^{c}, X^{s}$ need to be determined, 
except for the $D^{s}$-mode -- existing in the presence of external torques -- 
which is fixed by the torque balance:
in the general case, where an external torque is present,
we have (in the  co-moving frame)
\begin{equation}
D^{s}  =m\,,\,\,
m =\frac{M_h}{m_{drive}}\,,
%m =\frac{M_h}{SR\kappa l_{0}^{2}/2}\,,
\label{torque_Ds_relation}
\end{equation}
where $m$ is the non-dimensional torque. $M_h$ is the hydrodynamic resistance torque. 
If any external torque is neglected, one simply has $D^s=0$.

We exemplify the method for the equation for $C_{2}$. One inserts the mode ansatz into
$\omega\frac{\partial}{\partial\theta}C_{2}  =+k_{cut}D-k_{-1}C_{2}$,
to obtain
\begin{align*}
&\omega\frac{\partial}{\partial\theta}\left(C_{2}^{0}+C_{2}^{c}\cos\theta+C_{2}^{s}\sin\theta\right) 
% =&+k_{cut}\left(D^{0}+D^{c}\cos\theta+\underbrace{D^{s}\sin\theta}\right)\nonumber\\
%&-k_{-1}\left(C_{2}^{0}+C_{2}^{c}\cos\theta+C_{2}^{s}\sin\theta\right)\,,\\
%\rightarrow
=\omega\left(-C_{2}^{c}\sin\theta+C_{2}^{s}\cos\theta\right)  \nonumber\\
=&+k_{cut}\left(D^{0}+D^{c}\cos\theta+{D^{s}\sin\theta}\right)
-k_{-1}\left(C_{2}^{0}+C_{2}^{c}\cos\theta+C_{2}^{s}\sin\theta\right)\,.
\end{align*}
Now we project on the modes by performing integrations like
$\int d\theta$, $\int \sin\theta\,d\theta$ and $\int \cos\theta\,d\theta$.
This yields three conditions -- for the constant, sin- and cos-mode, respectively: 
\begin{align*}
0 & =k_{cut}D^{0}-k_{-1}C_{2}^{0}\,,\\
-\omega C_{2}^{c} & =-k_{-1}C_{2}^{s}+{k_{cut}D^{s}}\,,\\
\omega C_{2}^{s} & =k_{cut}D^{c}-k_{-1}C_{2}^{c}\,.
\end{align*}
Note that terms $\propto D^s$ are only present in the case of an external torque.
The equations can be solved and simplified.
Proceeding the same way for all equations, one gets expressions
for $B_1, C_1, C_2$ as functions of $D^0, D^c, D^s$
\begin{eqnarray}	
B_{1}^{0} & =&\frac{k_{-2}+k_{cut}}{k_{2}^{0}\left(1-\frac{\alpha^{2}}{2}\right)}\left(D^{0}-\frac{\alpha}{2}D^{c}\right){-\frac{\alpha}{2}\frac{\omega}{k_{2}^{0}\left(1-\frac{\alpha^{2}}{2}\right)}D^{s}}\,,\nonumber\\
B_{1}^{s} & =&-\frac{\omega}{k_{2}^{0}}D^{c}+{\frac{k_{-2}+k_{cut}}{k_{2}^{0}}D^{s}}\,,\nonumber\\
B_{1}^{c} & =&\frac{k_{-2}+k_{cut}}{k_{2}^{0}\left(1-\frac{\alpha^{2}}{2}\right)}\left(D^{c}-\alpha D^{0}\right)+{\frac{\omega}{k_{2}^{0}\left(1-\frac{\alpha^{2}}{2}\right)}D^{s}}\,,
\end{eqnarray}
\begin{eqnarray}
C_{1}^{0} & =&\frac{k_{cut}}{k_{-1}}B_{1}^{0}+C_{2}^{0}\,,\nonumber\\
C_{1}^{s} & =&\frac{k_{-1}\left(k_{cut}B_{1}^{s}+k_{-1}C_{2}^{s}\right)
	+\omega\left(k_{cut}B_{1}^{c}+k_{-1}C_{2}^{c}\right)}{k_{-1}^{2}+\omega^{2}}\,,\nonumber\\
C_{1}^{c} & =&\frac{k_{-1}\left(k_{cut}B_{1}^{c}+k_{-1}C_{2}^{c}\right)
	-\omega\left(k_{cut}B_{1}^{s}+k_{-1}C_{2}^{s}\right)}{\left(k_{-1}^{2}+\omega^{2}\right)}\,,
\end{eqnarray}	
\begin{eqnarray}	
C_{2}^{0} & =&\frac{k_{cut}}{k_{-1}}D^{0}\,,\nonumber\\
C_{2}^{s} & =&\frac{\omega k_{cut}}{k_{-1}^{2}+\omega^{2}}D^{c}+{\frac{k_{-1}k_{cut}}{k_{-1}^{2}+\omega^{2}}D^{s}}\,,\nonumber\\
C_{2}^{c} & =&\frac{k_{cut}k_{-1}}{k_{-1}^{2}+\omega^{2}}D^{c}-{\frac{k_{cut}\omega}{k_{-1}^{2}+\omega^{2}}D^{s}}\,,
\end{eqnarray}	
and three additional equations 
\begin{align}
\label{add1} 0  =&k_{1}G-k_{1}GC_{1}^{0}-k_{1}GC_{2}^{0}+(k_{-2}-k_{1}G)D^{0}\nonumber\\
 &-\left(k_{1}G+k_{-1}+k_{2}^{0}+k_{cut}\right)B_{1}^{0}
+k_{2}^{0}\alpha\frac{1}{2}B_{1}^{c}\,,\\
-\omega B_{1}^{c}  =&-k_{1}GC_{1}^{s}-k_{1}GC_{2}^{s}-\left(k_{1}G+k_{-1}+k_{2}^{0}+k_{cut}\right)B_{1}^{s}\nonumber\\
&+{(k_{-2}-k_{1}G)D^{s}}\,,  \label{add2} \\
\omega B_{1}^{s}  =&-k_{1}GC_{1}^{c}-k_{1}GC_{2}^{c}+(k_{-2}-k_{1}G)D^{c}\nonumber\\
&-\left(k_{1}G+k_{-1}+k_{2}^{0}+k_{cut}\right)B_{1}^{c}+k_{2}^{0}\alpha B_{1}^{0}\,, \label{add3}
\end{align}
for the three unknowns $D^{0}$, $D^{c}$,
and $\omega$ -- note that $D^s$ is not an unknown, since it is fixed by $m$ (or zero in absence of external torque).

\noindent {\bf Approximation of fast binding rate $k_2$.}
To be able to proceed, we now assume  that the binding of the second ``leg'' of a 
glycan dumbbell that is already attached to the virus, $k_2$, 
is faster than all other processes.
More specifically, we assume
$k_{2}^{0}\gg k_{others}\,,\,\omega$ while $k_{others}B_{1}\rightarrow0,\omega B_{1}\rightarrow0$
and only $B_{1}k_{2}^{0}\sim O\left(1\right)$ stays finite.

The effect is that we can drop all terms with $k_{others}B_{1}$,
express everything in terms of $D^0, D^c, D^s$ 
and get
$C_{1}^{0} =\frac{k_{cut}}{k_{-1}}D^{0}=C_{2}^{0}$
and
\begin{eqnarray}
C_{1}^{s}&=&\frac{k_{-1}k_{cut}}{\left(k_{-1}^{2}+\omega^{2}\right)^{2}}\left(2\omega k_{-1}D^{c}
	+{\left(k_{-1}^{2}-\omega^{2}\right)D^{s}}\right),\nonumber\\
C_{1}^{c}&=&\frac{k_{-1}k_{cut}}{\left(k_{-1}^{2}+\omega^{2}\right)^{2}}
	\left(\left(k_{-1}^{2}-\omega^{2}\right)D^{c}-{2\omega k_{-1} D^{s}}\right),
%C_{2}^{0} &=&\frac{k_{cut}}{k_{-1}}D^{0}\,,\nonumber\\
\label{solved_eqs}
\end{eqnarray}
while the other expressions for
$C_{2}^{0}$,  $C_{2}^{s}$, $C_{2}^{c}$, $B_{1}^{0}$, $B_{1}^{s}$ 
and $B_{1}^{c}$
stay the same. Eqs.~(\ref{add1})-(\ref{add3}) simplify to
\begin{align}
0 =&k_{1}G\left(1-C_{1}^{0}-C_{2}^{0}\right)+(k_{-2}-k_{1}G)D^{0}-k_{2}^{0}B_{1}^{0}+k_{2}^{0}\alpha\frac{1}{2}B_{1}^{c}\,,
\label{eq1_D}
\\
0 =&-k_{1}G\left(C_{1}^{s}+C_{2}^{s}\right)-k_{2}^{0}B_{1}^{s}+{(k_{-2}-k_{1}G)D^{s}}\,,
\\
0 =&-k_{1}G\left(C_{1}^{c}+C_{2}^{c}\right)+(k_{-2}-k_{1}G)D^{c}-k_{2}^{0}B_{1}^{c}+k_{2}^{0}\alpha B_{1}^{0}\,.
\label{eq3_D}
\end{align}
From Eqs.~(\ref{solved_eqs}), we can now extract
\[
\left(C_{1}^{s}+C_{2}^{s}\right)=\frac{\omega k_{cut}\left(3k_{-1}^{2}+\omega^{2}\right)}{\left(k_{-1}^{2}
+\omega^{2}\right)^{2}}D^{c}+{\frac{2k_{-1}^{3}k_{cut}}{\left(k_{-1}^{2}+\omega^{2}\right)^{2}}D^{s}}
\]
and
\[
\left(C_{1}^{c}+C_{2}^{c}\right)=\frac{2k_{-1}^{3}k_{cut}}{\left(k_{-1}^{2}+\omega^{2}\right)^{2}}D^{c}
-{\frac{k_{cut}\omega\left(3k_{-1}^{2}+\omega^{2}\right)}{\left(k_{-1}^{2}+\omega^{2}\right)^{2}}D^{s}}
\]
and inserting all known fields into Eqs.~(\ref{eq1_D})-(\ref{eq3_D})
results in three closed equations for $D^{0}$, $D^{c}$ and $\omega$
for given $D^s$:
%1st:
\begin{align}
0 =&k_{1}G+\left(k_{-2}-k_{1}G-\frac{2k_{cut}k_{1}G}{k_{-1}}-\frac{\left(k_{-2}+k_{cut}\right)\left(1+\frac{\alpha^{2}}{2}\right)}{\left(1-\frac{\alpha^{2}}{2}\right)}\right)D^{0}\nonumber\\
  &+\alpha\frac{k_{-2}+k_{cut}}{\left(1-\frac{\alpha^{2}}{2}\right)}D^{c}+{\frac{\alpha\omega}{\left(1-\frac{\alpha^{2}}{2}\right)}D^{s}},
 \label{eq1_Dfin}\\ 
0=&\left(\omega-\frac{k_{1}Gk_{cut}\left(3k_{-1}^{2}+\omega^{2}\right)\omega}{\left(k_{-1}^{2}+\omega^{2}\right)^{2}}\right)D^{c}\nonumber\\
&-{\left(\frac{k_{1}G2k_{-1}^{3}k_{cut}}{\left(k_{-1}^{2}+\omega^{2}\right)^{2}}+k_{1}G+k_{cut}\right)D^{s}},
\label{eq2_Dfin}\\
0 =&\left(-\frac{k_{1}G2k_{-1}^{3}k_{cut}}{\left(k_{-1}^{2}+\omega^{2}\right)^{2}}+k_{-2}-k_{1}G-\frac{\left(k_{-2}+k_{cut}\right)\left(1+\frac{\alpha^{2}}{2}\right)}{\left(1-\frac{\alpha^{2}}{2}\right)}\right)D^{c}\nonumber\\
  &+2\alpha\frac{k_{-2}+k_{cut}}{\left(1-\frac{\alpha^{2}}{2}\right)}D^{0}
  	+{\left(\frac{k_{1}Gk_{cut}\omega\left(3k_{-1}^{2}+\omega^{2}\right)}{\left(k_{-1}^{2}+\omega^{2}\right)^{2}}-\frac{\omega\left(\frac{\alpha^{2}}{2}+1\right)}{\left(1-\frac{\alpha^{2}}{2}\right)}\right)D^{s}}.
 \label{eq3_Dfin}
\end{align}

\noindent {\bf Case of negligible external torque/friction.}
The simplest result can be obtained, if one assumes that the external torque
-- which is due to hydrodynamic friction, and for such a small object as a 100 nm 
thin virus should be very small -- is negligible.
$m=0$ then immediately implies $D^s=0$, cf.~Eq.~(\ref{torque_Ds_relation}).

In this case, from Eq.~(\ref{eq2_Dfin}), we immediately get the following
instructive result:
\begin{eqnarray}
0&=&\omega\left(1-\frac{k_{1}Gk_{cut}\left(3k_{-1}^{2}+\omega^{2}\right)}{\left(k_{-1}^{2}+\omega^{2}\right)^{2}}\right)D^{c}\,.
\label{mot_rel_wo_m}
\end{eqnarray}
First of all, $\omega=0$  -- i.e.~no rotation at all -- is always a solution, as it should.
The term in brackets could yield non-trivial solutions, but only in case that
$D^c\neq0$. This also makes sense, since otherwise $D$ would be constant
and no torque is created.   
In fact, for $\alpha\neq0$ one has $D^c\neq0$ -- which can be seen from 
Eqs.~(\ref{eq1_Dfin}) and (\ref{eq3_Dfin}).
There is one more condition: $k_{-1}>0$ has also to be finite.
Otherwise one can show that $D(\omega)=0$, which reflects 
a ``blocking effect'': all initially existing double bound states were cut 
and the cut glycans never can detach.

In total, this yields (in case $\alpha, k_{-1}\neq0$)
three possible solutions for the angular velocity $\omega$:
\begin{equation}
\omega=0\,\,\,\,{\rm or}\,\,\,\,\,\omega^{2} =\frac{k_{1}Gk_{cut}}{2}\left(1+\sqrt{1+\frac{8k_{-1}^{2}}{k_{1}Gk_{cut}}}\right)-k_{-1}^{2},
\end{equation}
as given in the main text.
The latter yields two symmetric branches,
\begin{equation}
\omega = \pm\sqrt{  \frac{k_{1}Gk_{cut}}{2}\left(1+\sqrt{1+\frac{8k_{-1}^{2}}{k_{1}Gk_{cut}}}\right)-k_{-1}^{2} }\,\,.
\end{equation}
This is the pitchfork bifurcation explained in the main manuscript:
for sufficiently large cutting rate $k_{cut}$, 
namely if 
\begin{equation}
k_{cut}>k_{cut}^{c,0}  =\frac{k_{-1}^{2}}{3k_{1}G}
\end{equation} 
such that the radicand becomes positive,
two new branches with finite angular velocity
emerge from the trivial branch $\omega=0$.
The turning direction is spontaneously chosen, depending
on the initial conditions.

\noindent{\bf General case and obtaining the motor relation.}
To treat the general case, we  consider the limit of small $\alpha$,
i.e.~small variations of $k_2(\theta)$ with $\theta$.
We expand Eqs.~(\ref{eq1_Dfin})-(\ref{eq3_Dfin})
to linear order in $\alpha$, and 
notice that a certain symmetry emerges:
\begin{align}
0  =&k_{1}G-\left(k_{1}G+\frac{2k_{cut}k_{1}G}{k_{-1}}+k_{cut}\right)D^{0}\nonumber\\
&+\alpha\left(k_{-2}+k_{cut}\right)D^{c}+{\alpha\omega D^{s}}\,,
\label{eq1D_sma}\\
0=&
u(\omega)D^c
%-\underbrace{\left(\frac{k_{1}G2k_{-1}^{3}k_{cut}}{\left(k_{-1}^{2}+\omega^{2}\right)^{2}}+k_{1}G+k_{cut}\right)}_{v}D^{s}\,,\quad\quad\quad
-v(\omega)D^s\,,
\label{eq2D_sma}\\
0=&-v(\omega)D^c-u(\omega)D^s+2\alpha\left(k_{-2}+k_{cut}\right)D^{0},
\label{eq3D_sma} 
\end{align}
where we introduced the quantities
\begin{align}
u(\omega)  = & \omega\left(1-\frac{k_{1}Gk_{cut}\left(3k_{-1}^{2}+\omega^{2}\right)}{\left(k_{-1}^{2}+\omega^{2}\right)^{2}}\right)\,,\nonumber \\
v(\omega)  = & \left(\frac{2k_{1}Gk_{-1}^{3}k_{cut}}{\left(k_{-1}^{2}+\omega^{2}\right)^{2}}+k_{1}G+k_{cut}\right)\,.
\label{u_and_v}
\end{align}
From (\ref{eq2D_sma}) we immediately get
$D^{c}=\frac{v}{u}D^{s}$,
inserting into (\ref{eq3D_sma}) yields
\begin{align*}
%2\alpha\left(k_{-2}+k_{cut}\right)D^{0} & =vD^{c}+uD^{s}=\left(\frac{v^{2}}{u}+u\right)D^{s}\\
%\rightarrow\quad 
D^{0} & =\frac{\frac{v^{2}}{u}+u}{2\alpha\left(k_{-2}+k_{cut}\right)}D^{s}\,.
\end{align*}
and inserting into  (\ref{eq1D_sma}), one obtains
\begin{eqnarray}
0 & =&k_{1}G-\left(k_{1}G+\frac{2k_{cut}k_{1}G}{k_{-1}}+k_{cut}\right)\frac{\frac{v^{2}}{u}+u}{2\alpha\left(k_{-2}+k_{cut}\right)}D^{s}\label{eq:-3-3-1}\nonumber\\
 & &+\alpha\left(k_{-2}+k_{cut}\right)\frac{v}{u}D^{s}+\alpha\omega D^{s}\,.\nonumber
\end{eqnarray}
%Multiplying with $\alpha$ and 
Dropping orders $O\left(\alpha^{2}\right)$
%\begin{align*}
%0 & =\alpha k_{1}G-\left(k_{1}G+\frac{2k_{cut}k_{1}G}{k_{-1}}+k_{cut}\right)\frac{\frac{v^{2}}{u}+u}{2\left(k_{-2}+k_{cut}\right)}D^{s}\\
%\end{align*}
we solve for
\begin{equation}
D^{s}=\frac{2\alpha k_{1}G\left(k_{-2}+k_{cut}\right)}{v_{0}}\frac{u}{v^{2}+u^{2}},
\label{eq:Motor-Rel-Ds}
\end{equation}
where $v_0=v(\omega=0)$.
Remembering the relation between the torque and $D^{s}$, Eq.~(\ref{torque_Ds_relation}), 
%(with $D^{s}$ given by the small $\alpha$ expansion), $\frac{M}{k_{B}T}=\frac{\alpha}{2}D_{s}$ 
we finally obtain the  \textbf{"motor relation"}
\begin{equation}
M_h=m_{drive}\frac{\alpha^2 k_{1}G\left(k_{-2}+k_{cut}\right)}{v_{0}}\frac{u(\omega)}{v^{2}(\omega)+u^{2}(\omega)}.
\label{eq:Motor-Rel}		
\end{equation}
It relates any external torque (here the hydrodynamic frictional torque $M_h$) 
to the kinetics of the rotation.\\

\noindent{\bf Dynamics under Stokes friction.}
Assuming the external torque is Stokes friction 
in a fluid of viscosity $\eta$, yields the explicit expression
\[
M_{h}=4\pi\eta R^{2}l_{0}\omega
\]
and
\[
\frac{4\pi\eta R^{2}l_{0}\omega}{m_{drive}}=\alpha^{2}\frac{k_{1}G\left(k_{-2}+k_{cut}\right)}{v_{0}}\frac{u}{v^{2}+u^{2}}\,.
\]
Considering that for small torques, $u$ is small as well
-- note that $u\left(\omega\right)=0$ was the relation determining $\omega$
in the case of no frictional torque, cf.~Eq.~(\ref{mot_rel_wo_m}) -- 
we drop the $u^{2}$-term in the denominator and
expand the r.h.s.~for $\omega^{2}$ small 
(close to the onset of motion). 
The leading order constant term $O\left(\omega^{0}\right)$
cancels on both sides for the motion to occur
\[
\frac{\left(1+\frac{2k_{cut}}{k_{-1}}+\frac{k_{cut}}{k_{1}G}\right)}{\left(k_{-2}+k_{cut}\right)}\frac{4\pi R^{2}l_{0}}{\alpha^{2}m_{drive}}\eta=\frac{\left(1-\frac{3k_{1}Gk_{cut}}{k_{-1}^{2}}\right)}{\left(\frac{2k_{1}Gk_{cut}}{k_{-1}}+\left(k_{1}G+k_{cut}\right)\right)^{2}}\,,
\]
where $k_{cut}$ is now the critical value for the onset of  motion. 
This condition can now be solved perturbatively, 
for small l.h.s., in the relevant activity parameter $k_{cut}$. 
The latter deviates from the case without frictional torque,
$k_{cut}^{c,0}$, like
$k_{cut}  =k_{cut}^{c,0}+c\eta$   %\,\,,\,\,\,k_{cut}^{c,0}  =\frac{k_{-1}^{2}}{3k_{1}G}
with some proportionality constant $c$. 
Inserting this expansion we get $c$ 
to lowest order in $\eta$. 
%\begin{align*}
%\frac{\left(1+\left(\frac{2}{k_{-1}}+\frac{1}{k_{1}G}\right)k_{cut}^{0}\right)}{\left(k_{-2}+k_{cut}^{0}\right)}\frac{4\pi R^{2}l_{0}}{\alpha^{2}k_{B}T} & =\frac{\frac{1}{k_{cut}^{0}}c}{\left(\frac{2k_{1}Gk_{cut}^{0}}{k_{-1}}+\left(k_{1}G+k_{cut}^{0}\right)\right)^{2}}\\
%c & =\frac{k_{cut}^{0}\left(k_{1}G+\left(\frac{2k_{1}G}{k_{-1}}+1\right)k_{cut}^{0}\right)^{3}}{k_{1}G\left(k_{-2}+k_{cut}^{0}\right)}\frac{4\pi R^{2}l_{0}}{\alpha^{2}k_{B}T}
%\end{align*}
The result for the finite, but small viscosity correction 
to the critical cutting rate %$k_{cut}^c$
is (inserting $m_{drive}=\frac{S}{2}l_0^2R\kappa$): 
\begin{align}
k_{cut}^c =& k_{cut}^{c,0}+k_{cut}^{c,1}\left(\eta\right)\quad{\rm with}\quad
k_{cut}^{c,0}=\frac{k_{-1}^{2}}{3k_{1}G},\nonumber\\
& k_{cut}^{c,1} =
\frac{k_{cut}^{0}\left(k_{1}G+\left(\frac{2k_{1}G}{k_{-1}}+1\right)k_{cut}^{0}\right)^{3}}{k_{1}G\left(k_{-2}+k_{cut}^{0}\right)}
\frac{8\pi R}{\alpha^{2}S l_0\kappa}\,\eta\,.
\label{eq:-10}
\end{align}

\subsection*{D\quad Details on helix formation}

In general a virus surface displays multiple interdigitated and 
mutually staggered sub-lattices of switchable
spike-pairs, as sketched in Fig.~\ref{fig1_2}d)  (pairs 1-2 and 3-4).
We assume here for simplicity just two sub-lattices, I and II.
 
In the following we make the intuitive assumption that the curvature vector 
is a smooth function along the contour $s$, regardless of how it is measured, 
over sub-lattice I or II. 
This in particular implies that one sub-lattice will experience the mean effective 
curvature of the other sub-lattice 
(averaged over the smallest characteristic length scale $\sim l$). 
Due to the smoothness, the curvature vector felt by sub-lattice II 
at some position $s$ can be formally expressed by the mean of curvature vector 
over the sub-lattice I in its close vicinity, i.e.~as
\begin{equation}
\vec{\kappa}_{II}(s)=\frac{1}{l}\int_{s-l/2}^{s+l/2}\vec{\kappa}_{I}\left(s'\right)ds'.
\end{equation}

For a constant modulus $\kappa_{I}$, and a slowly rotating curvature
vector on the sub-lattice I,
\[
\vec{\kappa}_{I}\left(s\right)=\kappa_{I}\left(\cos\left(\tau s\right)\mathbf{e}_{x}+\sin\left(\tau s\right)\mathbf{e}_{y}\right)
\]
with some small torsion rate $\tau\ll l^{-1}$, 
the effective curvature of sub-lattice-II can be evaluated to
\begin{eqnarray}
\kappa_{II} 
&=&\kappa_{I}\frac{1}{l}\left| \left(\frac{1}{\tau}\sin\left(\tau s\right)|_{-l/2}^{+l/2}\mathbf{e}_{x}-\frac{1}{\tau}\cos\left(\tau s\right)|_{-l/2}^{+l/2}\mathbf{e}_{y}\right)\right|\nonumber\\
&=&\kappa_{I}\left|\frac{2\sin\left(\tau\,\frac{l}{2}\right)}{l\tau}\right| 
\approx\kappa_{I}\left(1-\frac{l^{2}\,\tau^{2}}{24}+\frac{l^{4}\tau^{4}}{1920}\right)+\ldots
%O\left(l^{6}\tau^{6}\right)\,.
\end{eqnarray}
That is, the   curvature felt by lattice II from the one
at lattice I  becomes effectively reduced by an amount
$-\kappa_{I}\left(\frac{l^{2}\tau^{2}}{24}-\frac{l^{4}\tau^{4}}{1920}\right)$. 

The overall energy penalty per length $l$ can then be written as
the average over the two sub-lattices, 
with $K_{tor}$ given by Eq.~(\ref{K_tor_est}),
and evaluates to
\begin{align*}
E_{tot} & =\frac{1}{2}\left(E_{I}+E_{II}\right)
 =\frac{1}{2}\left(\frac{1}{2}K_{tor}\left(\delta\kappa_{I}\right)^{2}+\frac{1}{2}K_{tor}\left(\delta\kappa_{II}\right)^{2}\right)\\
% & =\frac{1}{4}K_{tor}\left(\kappa_{tor}-\kappa_{I}\right)^{2}+\frac{1}{4}K_{tor}\left(\kappa_{tor}-\kappa_{I}\left(\frac{2\sin\left(\tau\,\frac{l}{2}\right)}{l\tau}\right)\right)^{2}\\
 &  \simeq\frac{1}{4}K_{tor}\left(2\left(\delta\kappa_{I}\right)^{2}+\kappa_{I}^{2}\left(\frac{l^{2}\,\tau^{2}}{24}\right)^{2}+2\left(\delta\kappa_{I}\right)\kappa_{I}\frac{l^{2}\,\tau^{2}}{24}\right)
 +\ldots
 %+O\left(\delta\kappa_{I}\kappa_{I}\left(\tau l\right)^{4}\right)
\end{align*}
with $\delta\kappa_{i}=\kappa_{tor}-\kappa_{i}$.

Dropping the index I and rearranging,
\begin{equation}
E_{tot}=\underbrace{\frac{K_{tor}}{2}\left(\kappa_{tor}-\kappa\right)^{2}}_{\text{Bending}}+\underbrace{\frac{1}{4}\frac{K_{tor}}{24^{2}}\kappa^{2}\left(l\tau\right)^{4}+\frac{K_{tor}}{2}\cdot\left(\kappa_{tor}-\kappa\right)\cdot\kappa\cdot\frac{\left(l\tau\right)^{2}}{24}}_{\text{Torsion-bend coupling}}\,,\nonumber
\end{equation}
we see that there is a classical bending energy term followed by an
unusual torsion-bend coupling. Overall the energy is positive definite
and has a minimum for $\kappa=\kappa_{tor}$ and $\tau=0$ as seen
from the form
\begin{equation}
E_{tot}=\frac{1}{4}K_{tor}\left(\kappa_{tor}-\kappa\right)^{2}+\frac{1}{4}K_{tor}\left(\left(\kappa_{tor}-\kappa\right)+\kappa\frac{l^{2}\,\tau^{2}}{24}\right)^{2}\label{eq: Kappa-Tau Energy}
\end{equation}
given in the main manuscript.

The torsion-bend  coupling suggests an unusual mechanics: 
(i) For optimal curvature $\kappa=\kappa_{tor}$ the torsional term
has a weak quartic stiffness $O\left(\tau^{4}\right)$ indicating
some flexibility in this degree of freedom.
(ii) For curvature larger than optimal, i.e.~for $\kappa_{tor}-\kappa<0$,
a torsion (of any sign) is favorable, given by
\begin{equation}
\tau_{\pm}=\pm\frac{1}{l}\sqrt{24}\sqrt{1-\frac{\kappa_{tor}}{\kappa}}\,,
\end{equation}
as already stated in the main text.
(iii) For curvature smaller than optimal, $\kappa_{tor}-\kappa>0$, the
torsion experiences a quadratic stiffness term 
$\frac{K_{tor}\left(\kappa_{tor}-\kappa\right)\cdot\kappa l^{2}}{48}\cdot\tau^{2}$
in leading order, forcing it strongly towards the $\tau=0$ state.

Overall this indicates some highly unusual spatial dynamics of the toro-elastic
state, very much different from usual semiflexible filaments. 
Note that case (ii), i.e.~larger curvature than $\kappa_{tor}$, is the one that should be 
relevant for a virus in a mucus environment, where glycan ligands cause additional 
tensile stresses, leading to increased curvatures.

\section*{Author Contributions}
F.Z.~and I.M.K.~developed the model. 
F.Z., K.G.D. and I.M.K.~performed the calculations.
All authors wrote the paper.

\section*{Conflicts of interest}
There are no conflicts to declare.
%In accordance with our policy on \href{https://www.rsc.org/journals-books-databases/journal-authors-reviewers/author-responsibilities/#code-of-conduct}{Conflicts of interest} please ensure that a conflicts of interest statement is included in your manuscript here.  Please note that this statement is required for all submitted manuscripts.  If no conflicts exist, please state that ``There are no conflicts to declare''.

\section*{Acknowledgements}
I.M.K. thanks Jens-Uwe Sommer for continuous discussion and support.  
%The Acknowledgements come at the end of an article after Conflicts of interest and before the Notes and references.

%%%END OF MAIN TEXT%%%


\balance


%%%REFERENCES%%%
\bibliography{virus.bib} %You need to replace "rsc" on this line with the name of your .bib file
\bibliographystyle{rsc} %the RSC's .bst file
%\nocite{*}%%%%%%%%%%%% delete later

\end{document}